\documentclass[adraft,creativecommons,noreriv]{eptcs}
\providecommand{\event}{ML 2016} 

\usepackage{hyphenat}
\usepackage{mdwtab}
\usepackage{mdwlist}
\usepackage{amsmath}
\usepackage{amssymb}
\usepackage{comment}
\usepackage{url}
\usepackage{graphicx}
\usepackage{prooftree}

\usepackage{listings}

\lstdefinestyle{sOcaml}{language=[Objective]Caml,
  morekeywords={effect,operation,handle},
  literate={+}{{$+\:$}}1 {/}{{$/$}}1 
           {=}{{$=$}}1
           {++}{{+\!+}}2
           {>}{{$>$}}1 {<}{{$<$}}1
           {<>}{$\not=\ $}1
           {->}{{$\rightarrow$}}2 {>=}{{$\geq$}}2 {<-}{{$\leftarrow$}}2
           {<=}{{$\leq$}}2
           {=>}{{$\Rightarrow$}}2
           {+->}{{$\hookrightarrow$}}2
           {==>}{{$\mapsto$}}2
           {fn}{$\lambda$}1
           {|}{{$\mid$}}1
           {'a}{$\alpha$}1
           {+'a}{$\textrm{+}\alpha$}1
           {'b}{$\beta$}1
           {+'b}{$\textrm{+}\beta$}1
           {'c}{$\gamma$}1
           {'e}{$\epsilon$}1
           {'w}{$\omega$}1
           {'w.}{$\forall\omega.\ $}2
           {t1}{t$_1$}2
           {t2}{t$_2$}2
           {\ .\ }{$\;\circ\;$}1
           {EV}{$\mathcal{E}$}1
           {===}{$\equiv$}1
           {\#}{{\textrm{\#}}}1
           {TRB}{\mbox{\ensuremath\lceil}}1
           {TRE}{\mbox{\ensuremath\rceil}}1
           {...}{\ldots}2
           {forall}{{$\forall$}}1
           {\#\#\#}{{$\leadsto$}}3
}
\lstnewenvironment{code}{\lstset{basicstyle={\sffamily\footnotesize}}}{}
\lstnewenvironment{codeE}{\lstset{basicstyle={\sffamily\footnotesize},frame=L}}{}

\lstMakeShortInline[columns=fullflexible]|

\newtheorem{proposition}{Proposition} 

\newcommand{\aside}[1]{\ignorespaces}

\begin{document}

\title{Eff Directly in OCaml}
\author{Oleg Kiselyov
\institute{Tohoku University, Japan}
\email{oleg@okmij.org}
\and
KC Sivaramakrishnan
\institute{University of Cambridge, UK}
\email{sk826@cam.ac.uk}
}
\def\authorrunning{Oleg Kiselyov \& KC Sivaramakrishnan}
\def\titlerunning{Eff Directly in OCaml}
\maketitle

\begin{abstract}
The language Eff is an OCaml-like language serving as a prototype
implementation of the theory of algebraic effects, intended for
experimentation with algebraic effects on a large scale.

We present the embedding of Eff into OCaml, using the library of
delimited continuations or the multicore OCaml branch.  We demonstrate
the correctness of the embedding denotationally, relying on the
tagless-final--style interpreter-based denotational semantics,
including the novel, direct denotational semantics of multi-prompt delimited
control.  The embedding is systematic, lightweight, performant and
supports even higher-order, `dynamic' effects with their polymorphism.
OCaml thus may be regarded as another implementation of Eff,
broadening the scope and appeal of that language.
\end{abstract}

\section{Introduction}

Algebraic effects \cite{PlotkinP03,Plotkin-handlers} are becoming
a more and more popular approach for expressing and composing
computational effects. There are implementations of algebraic effects
in Haskell \cite{kammar-handlersinaction,freer}, Idris
\cite{Brady-effects}, OCaml \cite{ocaml-effect,kammar-handlersinaction}, Koka
\cite{Leijen2017koka},
Scala\footnote{\url{https://github.com/atnos-org/eff},
\url{https://github.com/m50d/paperdoll}, among others},
Javascript\footnote{\url{https://www.humblespark.com/blog/extensible-effects-in-node-part-1}},
PureScript\footnote{\url{http://purescript.org}}, and other
languages. The most direct embodiment of algebraic effect theory is
the language Eff\footnote{\url{http://www.eff-lang.org/}} ``built to
test the mathematical ideas of algebraic effects in practice''. It is
an OCaml-like language with the native facilities (syntax, type
system) to declare effects and their handlers \cite{eff-language}. It
is currently implemented as an interpreter, with an optimizing compiler to OCaml
in the works.

Rather than compile Eff to OCaml, we \emph{embed} it. After all, save
for algebraic effects, Eff truly is a subset of OCaml and can be 
interpreted or compiled as a regular OCaml code. Only
effect declaration, invocation and handling need translation, which
is local and straightforward. It relies on the
library of delimited control delimcc \cite{caml-shift} or else
the Multicore OCaml branch \cite{ocaml-effect}. 
The embedding, in effect, becomes a set of OCaml
idioms for effectful programming with the almost exact look-and-feel
of Eff\footnote{While writing this paper we have implemented delimited
control in yet another way, in pure OCaml: see Core delimcc in 
\S\ref{s:sem-delimcc}. Although developed for the formalization of the
Eff translation, it may be used in real programs~-- provided the code is written
in a particular stylized way as shown in \S\ref{s:sem-delimcc}.
In contrast, the original delimcc and Multicore OCaml can be used 
with the existing OCaml code as it is. Hence they let Eff be embedded rather
than compiled into OCaml.
}.

Our second contribution is the realization that
so-called `dynamic effects', or handled `resources', of Eff 3.1
(epitomized by familiar reference cells, which can be created in any
number and hold values of any type) is not a separate language feature.
Rather, the dynamic creation of effects is but another effect,
and hence is already supported by our implementation of ordinary effects
and requires no special syntax or semantics.

As a side contribution, we show the correctness of our embedding of
Eff in OCaml denotationally, relying on the ``tagless-final'' style 
\cite{carette-finally,tagless-final-oxford} of
interpreter-based denotational semantics (discussed in more
detail in \S\ref{s:what-is-denot}). We also demonstrate the novel
denotational semantics of multi-prompt delimited control that does not
rely on continuation-passing-style (and is, hence, direct).

The structure of the paper is as follows. First we informally
introduce Eff on a simple example.  \S\ref{s:eff-delimcc} then
demonstrates our translation to OCaml using the delimcc library,
putting Eff and the corresponding OCaml code
side-by-side. \S\ref{s:eff-multicore} shows how the embedding works in
multicore OCaml with its `native effects'. \S\ref{s:formal} gives the
formal, denotational treatment, reminding the denotational semantics
of Eff; describing the novel direct denotational semantics of
multi-prompt delimited control; then presenting the translation
precisely; and arguing that it is meaning-preserving.  We describe the
translation of the dynamic effects into OCaml in \S\ref{s:higher}.
The empirical \S\ref{s:eval} evaluates the performance of our
implementation of Eff comparing it with the Eff's own optimizing
compiler.  Related work is reviewed in \S\ref{s:related}.  We then
conclude and summarize the research program inspired by our Eff
embedding.

The source code of all our examples and benchmarks is available at
\url{http://okmij.org/ftp/continuations/Eff/}.

\section{Eff in Itself and OCaml}

We illustrate the Eff embedding on the running example, juxtaposing
the Eff code with the corresponding OCaml. We thus demonstrate both
the simplicity of the translation and the way to do Eff-like effects
in idiomatic OCaml.

\subsection{A taste of Eff}
\label{s:eff-intro}

An effect in Eff has to be declared first\footnote{Eff code is marked
with double vertical lines to distinguish it from OCaml.}:
\begin{codeE}
type 'a nondet = effect
  operation fail   : unit -> empty
  operation choose : ('a * 'a) -> 'a
end
\end{codeE}
Our running effect is thus familiar non-determinism.  The
declaration introduces only the \emph{names} of effect operations~--
the failure and the non-deterministic choice between two alternatives~--
and
their types. The semantics is to be defined by a handler later on.
All effect invocations uniformly take an argument (even if it is dummy
|()|) and promise to produce a value (even if of the type |empty|, of
which no values exists; the program hence cannot continue after a
|failure|). The declaration is parameterized by the type |'a| of the
values to non-deterministically choose from.
(The parameterization can be avoided, if we rather gave |choose|
the type |unit->bool| or the (first-class) polymorphic type
|forall 'a. 'a * 'a -> 'a|.)

Next we ``instantiate the effect signature'', as Eff puts it:
\begin{codeE}
let r = new nondet
\end{codeE}
One may think of an instance |r| as part of the name for effect
operations: the signature |nondet| defines the common
interface. Different parts of a program may independently use
non-determinism if each creates an instance for its own use. Unlike
the effect declaration, which is static, one may create arbitrarily
many instances at run-time.

We can now write the sample non-deterministic Eff code:
\begin{codeE}
let f () =
  let x = r#choose ("a", "b") in
  print_string x ;
  let y = r#choose ("c", "d") in
  print_string y
\end{codeE}
The computation (using the Eff terminology \cite{eff-language})
|r#choose ("a", "b")| invokes the effect |choose| on instance |r|,
passing the pair of strings |"a"| and |"b"| as parameters. Indeed the
instance feels like a part of the name for an effect operation.  The
name of the effect hints that we wish to choose a string from the
pair. Strictly speaking however, |choose| does not have any meaning
beyond signaling the intention of performing the `choose' effect,
whatever it may mean, on the pair of strings.

To run the sample code, we have to tell how to interpret the effect
actions |choose| and |fail|: so far, we have only defined their names
and types: the algebraic signature. It is the
interpreter of the actions, the handler, that infuses the action
operations with their meanings. For example, Eff may execute the sample
code by interpreting |choose| to follow up on both choices,
depth-first:
\begin{codeE}
let test1 = handle f () with
  | val x             -> x
  | r#choose (x, y) k -> k x ; k y
  | r#fail () _       -> ()
\end{codeE}
The |handle...with| form is deliberately made to look like 
the |try...with| form of OCaml~--
following the principle that algebraic effects are a
generalization of ordinary exceptions. The |fail| action is treated as
a genuine exception: if the computation |f ()| invokes  |fail ()|,
|test1| immediately returns with |()|.
When the computation |f ()| terminates with a value, the
|val x| branch of the handle form is evaluated, with |x| bound to
that value; |test1| hence returns the result of |f ()|
as it is\footnote{An astute reader must have noticed that this result
  must again be unit.}. The |choose| action is the proper effect:
when the evaluation of |f ()| comes across |r#choose ("a","b")|,
the evaluation is suspended and the |r#choose| clause of the handle
form above is executed, with |x| bound to |"a"|, |y| bound to |"b"|
and |k| bound to the continuation of |f ()| evaluation, up to the
handle. Since |k| is the delimited continuation, it acts as a
function, returning what the entire handle form would return
(again unit, in our case). Thus the semantics given to |r#choose| in
|test1| is first to choose the first component of the pair;
and after the computation with the choice is completed, choose the
second component. The |choose| effect hence acts as a \emph{resumable}
exception, familiar from Lisp. In our case, it is in fact resumed twice.
Executing |test1| prints |acdbcd|.

Just like the |try| forms, the handlers may nest: and here things
become more interesting. First of all, distinct effects~-- or distinct
instances of the same effect~-- act independently, unaware of each
other. For example, it is rather straightforward to see
that the following code (where the |i1| and |i2| handlers make the
choice slightly differently)\footnote{
The effects \textsf{i1\#choose} and \textsf{i2\#choose} can also be
handled by the same handler: we touch on so-called multiple-effect
handlers in \S\ref{s:core-eff}.}
\begin{codeE}
let test2 =
 let i1 = new nondet in
 let i2 = new nondet in
 handle
   handle
     let x = i1#choose ("a", "b") in
     print_string x ;
     let y = i2#choose ("c", "d") in
     print_string y
   with
   | val () -> print_string ";"
   | i2#choose (x,y) k -> k x; k y
 with
 | val x -> x
 | i1#choose (x,y) k  -> k y; k x
\end{codeE}
prints |bc;d;ac;d;| The reader may try to work out the result when
the inner handler handles the |i1| instance and the outer one |i2|.

One may nest handle forms even for the same effect instance. To
confidently predict the behavior in that case one really needs the
formal semantics, overviewed in \S\ref{s:sem-eff}. First, the effect
handling code may itself invoke effects, including the very same
effect:
\begin{codeE}
let testn1 =
 handle
  handle
   let x = r#choose ("a", "b") in
   print_string x
  with
   | val () -> print_string ";"
   | r#choose (x,y) k -> k (r#choose(x,y))
 with
  | val x -> x
  | r#choose (x,y) k  -> k y; k x
\end{codeE}
The effect ``re-raised'' by the inner handler is then dealt with by an
outer handler. In |testn1| hence the inner handler simply relays the
|choose| action to the outer one. The code prints |b;a;|

A handler does not have to handle all actions of a signature. The
unhandled ones are quietly ``re-raised'' (again, similar to ordinary
exceptions):
\begin{codeE}
let testn2 =
 handle
  handle
   let x = r#choose ("a", "b") in
   print_string x;
   (match r#fail () with)
  with
   | val ()      -> print_string ";"
   | r#fail () _ -> print_string "!"
 with
  | val x -> x
  | r#choose (x,y) k  -> k y; k x
\end{codeE}
The code prints |b!a!|.
The main computation does both |fail| and |choose| effects; the inner
handler deals only with |fail|, letting |choose| propagate to the
outer one. An unhandled effect action is a run-time error.
The suspicious |(match r#fail () with)| expression does a case
analysis on the |empty| type. There are no values of that type and hence
no cases are needed.

Eff has another syntax for handling effects in an expression |e|: 
|with eh handle e|, where |eh| should evaluate to a value of the handler
type. Such values are created by the |handler| form: whereas
|handle...with| is meant to evoke |try...with|, the |handler| form is
reminiscent of OCaml's |function|. Just as |function| creates a
function value from a collection of clauses pattern-matching on the
argument, |handler| creates a handler value from a collection of
clauses pattern-matching on the effect operation. An example should
make it clear: the following code re-writes the earlier |testn2| in
|with eh handle e| notation:
\begin{codeE}
let testn2' =
 let hinner = handler 
   | val ()      -> print_string ";"
   | r#fail () _ -> print_string "!"
 in
 let houter = handler
  | val x -> x
  | r#choose (x,y) k  -> k y; k x
 in
 with houter handle
  with hinner handle
   let x = r#choose ("a", "b") in
   print_string x;
   (match r#fail () with)
\end{codeE}
The |with eh handle e| and |handler| notation emphasizes that handlers are
first-class values in Eff, and may hence be assigned a denotation.
For this reason, the paper \cite{eff-language} uses the notation
exclusively~-- and so does Core Eff in \S\ref{s:core-eff}.

\subsection{Eff in OCaml}
\label{s:eff-delimcc}

We now demonstrate how the Eff examples from the previous section
can be represented in OCaml, using the library of delimited control
delimcc \cite{caml-shift}. We intentionally write the OCaml code to
look very similar to Eff, hence showing off the Eff idioms and
introducing the translation from Eff to OCaml intuitively. We make the
translation formal in \S\ref{s:formal}.

Before we begin, we declare two OCaml types:
\begin{code}
type empty
type 'e result = Val | Eff of 'e
\end{code}
The abstract type |empty| is meant to represent the empty type of Eff,
the type with no values\footnote{The fact that \textsf{empty} has no
  constructors does not mean it cannot have any: after all, the type
  is abstract. Defining truly an empty type in OCaml is quite a
  challenge, which will take us too much into the OCaml specifics.}.
The |result| type represents results of handled computations, or
the domain of results |R| from
\cite[\S4]{eff-language}, to be described in more detail in
\S\ref{s:formal}. It is indexed only by the type of effects but
not by the type of the normal computational result, as we shall
discuss in detail later in this section.

We now begin with our translation, juxtaposing Eff code with the
corresponding OCaml. Recall, an effect has to be declared
first\footnote{Again, the Eff code is marked
with double vertical lines to distinguish it from OCaml.}:
\begin{codeE}
type 'a nondet = effect
  operation fail   : unit -> empty
  operation choose : ('a * 'a) -> 'a
end
\end{codeE}
In OCaml, an Eff declaration is rendered as a data type declaration:
\begin{code}
type 'a nondet =
  | Fail   of unit    * (empty -> 'a nondet result)
  | Choose of ('a * 'a) * ('a  -> 'a nondet result)
\end{code}
that likewise defines the names of effect operations, the types of
their arguments and the type of the result after invoking the effect.
The translation pattern should be easy to see: each data type variant has
exactly two arguments, the latter is the continuation. The
attentive reader quickly recognizes the freer monad
\cite{freer}.

To make the translation correspond even closer to Eff, we define two
functions |choose| and |fail|, using the delimited control operator
|shift0| provided by the delimcc library\footnote{
Our \textsf{shift0} operator is the multi-prompt version of
\textsf{shift0} that was introduced as a variation of the more familiar
\textsf{shift} in \cite{danvy-functional}. The `body' of
\textsf{shift0} in the present paper is always a value, in which case
\textsf{shift0} is equivalent to \textsf{shift}, only slightly faster.}
\begin{code}
let choose p arg = shift0 p (fun k -> Eff (Choose (arg,k)))
(* val choose : 'a nondet result Delimcc.prompt -> 'a * 'a -> 'a = <fun> *)
let fail p arg   = shift0 p (fun k -> Eff (Fail (arg,k)))
(* val fail : 'a nondet result Delimcc.prompt -> unit -> empty = <fun> *)
\end{code}
The inferred types of these functions are shown in the comments. The
first argument |p| is a so-called prompt \cite{caml-shift}, the
control delimiter. The delimcc operation
\begin{code}
val push_prompt  : 'a prompt -> (unit -> 'a) -> 'a
\end{code}
runs the computation (given as a thunk in the second argument) having
established the control delimiter.
The operator |shift0 p (fun k -> body)| captures and removes the continuation
up to the dynamically closest occurrence of a |push_prompt p|
operation, for the same value of |p|. It then evaluates |body|.
The captured continuation is
packed into a closure bound to |k|. We formally describe the semantics of
|shift0| in \S\ref{s:sem-delimcc}; for now one may think of the above
|choose| and |fail| functions as throwing an `exception' |Eff|~--
the exception that may be `recovered from', or resumed, when
the closure |k| is invoked. We observe that the |fail| and
|choose| definitions look entirely regular and could have been mechanically
generated. The inferred types look almost like the types of the
corresponding |Eff| operations. For example,
our |choose| is quite like Eff's |r#choose|: it takes the
effect instance (prompt) and a pair of values and
(non-deterministically) returns one of them. Strictly speaking,
however, |choose| (just like |r#choose| in Eff) does hardly anything:
it merely captures the
continuation and packs it, along with the argument,
in the data structure, to be passed to the
effect handler. The handler does the choosing.

The ``instantiation of the effect signature''
\begin{codeE}
let r = new nondet
\end{codeE}
looks into OCaml as creating a new prompt
\begin{code}
let r = new_prompt ()
\end{code}
whose type, inferred from the use in the code below, is
|string nondet result prompt|. The type does look like the type of an
`instance' of the |nondet| effect. The created prompt can be passed as
the first argument to the |choose| and |fail| functions introduced
earlier.

We can now translate the sample Eff code that uses non-determinism
\begin{codeE}
let f () =
  let x = r#choose ("a", "b") in
  print_string x ;
  let y = r#choose ("c", "d") in
  print_string y
\end{codeE}
into OCaml as
\begin{code}
let f () =
  let x = choose r ("a","b") in
  print_string x ;
  let y = choose r ("c","d") in
  print_string y
\end{code}
The translation is almost literally copy-and-paste, with small
stylistic adjustments.
The effect instance |r| is passed to |choose| as the regular argument,
without any special |r#| syntax.

To run our sample Eff code or its OCaml translation we have to define
how to interpret the |choose| effects. In Eff, it was the job of the
handler. Recall:
\begin{codeE}
let test1 = handle f () with
  | val x             -> x
  | r#choose (x, y) k -> k x ; k y
  | r#fail () _       -> ()
\end{codeE}
The handler has two distinct parts: one defining the interpretation of
the result of |f ()| execution (the \textsf{val x} clause); the rest
deals with interpreting effect operations and resuming the
computation interrupted by these effects. (Or not resuming,
if the resumption, i.e., continuation bound to |k|, is not invoked:
see the |r#fail| clause). The form of the
handler expression almost makes it look as if a computation such as
|f ()| may end in two distinct ways: normally, yielding a value, or
by performing an effect operation. In the latter case, the result collects
the arguments passed to the effect operation plus the continuation to
resume the computation after the effect is handled. The
denotational semantics of Eff presented in \cite[\S4]{eff-language}
and reminded in \S\ref{s:sem-eff} gives computations exactly such
a denotation: a terminating computation is either a value or an
effect operation with its arguments and the continuation. Our
translation of Eff to OCaml takes such
denotation to heart, representing it by the |'e result| type.

At first glance, the |result| type should have been defined as
\begin{code}
type ('w,'e) result_putative = Val of 'w | Eff of 'e
\end{code}
with two parameters: |'e| being the type of the effect and
|'w| being the type of the normal result. The two type parameters look
independent, as expected. This type is the type of a handled
computation~-- and, hence, the result type of a resumption
(continuation) of this computation. The |nondet| effect,
whose operation carries such continuation, should, therefore,
have been defined as
\begin{code}
type ('w,'a) nondet_putative =
  | Fail   of unit    * (empty -> ('w,('w,'a) nondet_putative) result_putative)
  | Choose of ('a * 'a) * ('a  -> ('w,('w,'a) nondet_putative) result_putative)
\end{code}
We have no choice but to make |'w| also a parameter of the
|nondet_putative| lest the type variable |'w| be left unbound.
The effect type and the normal result type are not independent after
all. The surprising occurrence of |'w| in the effect type is not just
aesthetically disappointing. The effect instance (prompt) type also
becomes parameterized by |'w|. Therefore, if we use a
|nondet| effect instance in a computation that eventually produces
|int|, we cannot use the instance in a computation that
eventually produces |bool|. (Recall that prompt types cannot be
polymorphic: after all, delimited control can easily emulate mutable
state, with prompt playing the role of the reference cell
\cite{delimited-dyn-binding}.)

Strictly speaking, we need so-called answer-type polymorphism
\cite{asai-polymorphic}~-- which, however, cannot be added to OCaml
without extensive changes to its type system. Fortunately, it can
be cheaply, albeit underhandedly, emulated. For example, we can
`cast away' the normal result type with the help of the universal
type:
\begin{code}
type 'e result_v1 = Val of univ | Eff of 'e
\end{code}
The type of the handled computation is now parameterized solely by the
effect type; the troublesome answer-type dependence on |'w| is now gone.
The universal type can be emulated in OCaml in several ways; for
example, as |Obj.t|\footnote{See also \url{http://mlton.org/UniversalType}}.
 A safer way (in the sense that mistakes
in the emulation code lead to a
run-time OCaml exception rather than a segmentation fault)
is to carry the normal
computation result `out of band'. In which case, the handled
computation gets the simpler type
\begin{code}
type 'e result = Val | Eff of 'e
\end{code}
which was defined at the beginning of this section.
Such an out-of-band trick was earlier used in
\cite[\S5.2]{delimited-dyn-binding}, which also explains the need for the
polymorphism in more detail.

To carry the normal computation result out-of-band, we use a
reference cell:
\begin{code}
type 'a result_value = 'a option ref
let get_result : 'a result_value -> 'a = fun r ->
  match !r with
    Some x -> r := None; (* GC *) x
\end{code}
One is reminded of a similar trick of extracting the result of a
computation in continuation-passing style\footnote{If the
  continuation is given the type $\alpha\to\mathsf{empty}$ then the
  often heard `pass the identity continuation' is type-incorrect.}
which is often used in implementations of delimited control
(for example, \cite{caml-shift})\footnote{We could also have used a
  related trick: exceptions.}.
The reference cell |'a result_value|
is allocated and stored into in the following code\footnote{The
right-associative infix operator \textsf{@@} of low
precedence is application: \textsf{f @@ x + 1} is the same as
\textsf{f (x + 1)} but avoids the parentheses. The operator is the
analogue of \textsf{\char`$} in Haskell.}: 
\begin{code}
let handle_it:
    'a result prompt ->                     (* effect instance *)
    (unit -> 'w) ->                           (* expression *)
    ('w -> 'c) ->                              (* val clause *)
    (('a result -> 'c) -> 'a -> 'c) ->       (* oper clause *)
    'c =
  fun effectp exp valh oph ->
  let res = ref None in
  let rec loop : 'a result -> 'c = function
    | Val     -> valh (get_result res)
    | Eff eff -> oph loop eff
  in loop @@ push_prompt effectp @@ fun () -> (res := Some (exp ()); Val)
\end{code}
The expression to handle (given as a thunk |exp|) is run after setting
the prompt to delimit continuations captured by effect operations
(more precisely, by |shift0| underlying |choose|
and other effect operations). If the computation finishes, the value
is stored, for a brief moment, in the reference cell |res|, and then
extracted and passed to the normal termination handler |valh|. Seeing
how |handle_it| is actually used may answer the remaining questions about
it:
\begin{code}
let test1 = handle_it r f
  (fun x -> x) @@ fun loop -> function
    | Choose ((x,y),k) -> loop (k x); loop (k y)
    | Fail ((),_)      -> ()
\end{code}
The OCaml version of |test1| ends up very close to the Eff version.
We can see that |handle_it| receives the `effect instance' (the prompt
|r|), the thunk |f| of the computation to perform, and two handlers,
for the normal termination result (which is the identity in our case,
corresponding to the clause |val x -> x| in the Eff code)
and for handling the |'a nondet| operations, |Choose| and |Fail|. The only
notable distinction from Eff is how we resume the continuation:
we now write |loop (k x)| as compared to the
simple |k x| in Eff. As we shall see
in \S\ref{s:sem-eff}, even in Eff the resumption has the form of
invoking the captured expression continuation,
whose result is then fed into an auxiliary
recursive function, called |loop| here (and called |h| in
Fig.~\ref{f:eff-sem}). For convenience, Eff offers the user the already
composed resumption; the handlers receiving such composed resumption
are called deep.

The just outlined translation applies to the nested handlers as
is. For example, the |test2| code from \S\ref{s:eff-intro} is
translated into OCaml as follows:
\begin{code}
let test2 =
 let i1 = new_prompt () in
 let i2 = new_prompt () in
 handle_it i1 (fun () ->
   handle_it i2 (fun () ->
     let x = choose i1 ("a", "b") in
     print_string x ;
     let y = choose i2 ("c", "d") in
     print_string y)
   (fun () -> print_string ";") @@ fun loop -> function
   | Choose ((x,y),k) -> loop (k x); loop (k y)
 )
 (fun x -> x) @@ fun loop -> function
 | Choose ((x,y),k)  -> loop (k y); loop (k x)
\end{code}
Here, the inner normal termination handler is not
the identity: it performs the printing, just like the corresponding
Eff value handler |val () -> print_string ";"|.
The translation was done by copying-and-pasting of the Eff code and
doing a few slight modifications. The code runs and prints the same
result as the original Eff code. The other nested handling examples,
|testn1| and |testn2| of \S\ref{s:eff-intro} are translated in the
manner just outlined, and just as straightforwardly. We refer to the
source code for details.

\subsection{Eff in multicore OCaml}
\label{s:eff-multicore}

In this section, we describe the embedding of Eff in multicore OCaml. But first
we briefly describe the implementation of algebraic effects and handlers in
multicore OCaml.

\subsubsection{Algebraic effects in multicore OCaml}

Multicore OCaml~\cite{multicoreocaml} is an extension of OCaml with native support
for concurrency and parallelism. Concurrency in multicore OCaml is expressed
through algebraic effects and their handlers. We might declare the
non-determinism operations as:
\begin{code}
effect Fail   : empty
effect Choose : ('a * 'a) -> 'a
\end{code}
Unlike Eff, multicore OCaml does not provide the facility to define new effect
types. Indeed, the above declarations are simply syntactic sugar for extending
the built-in effect type with new operations:
\begin{code}
type _ eff += Fail   : empty eff
type _ eff += Choose : 'a * 'a -> 'a eff
\end{code}
The |test1|-like example (see \S\ref{s:eff-intro}) takes the
following form:
\begin{code}
let f () =
  let x = perform (Choose ("a","b")) in
  print_string x;
  let y = perform (Choose ("c";"d")) in
  print_string y
in
match f () with
| x -> x (* value clause *)
| effect Choose(x,_) k -> continue k x
| effect Fail _ -> ()
\end{code}
Effects are performed with the |perform| keyword. Multicore OCaml extends
OCaml's pattern matching syntax to double up as handlers when effect patterns
(patterns that begin with the keyword |effect|) are present. Unlike
the real |test1| however, this multicore OCaml example
always chooses the first component of the pair, for the reasons
detailed below. The continuation |k| is not
a closure and is resumed with
the |continue| keyword.
\aside{The handlers in OCaml are deep and the handlers wrap around
the continuation.} Just like ambient effects in OCaml, user-defined effects in
multicore OCaml have no type-level marker that decorates function types with
effects performed. An effect that is not handled by any handler in the current
stack raises a runtime exception.

Algebraic effects were developed in multicore OCaml primarily to
support concurrency; therefore, by default, the continuations are
one-shot and can be resumed at most once. This restriction is enforced
with dynamic checks, which raise an exception when a continuation is
resumed more than once. Pleasantly, this restriction allows multicore
OCaml to implement the continuations in \emph{direct-style}, by
creating a new heap-managed stack object for effect
handlers. Continuation capture is also cheap; capturing a continuation
only involves obtaining a reference to the underlying stack
object. Since the continuations are one-shot, there is no need for
copying the continuation object when resuming the continuation. For
OCaml, these direct-style continuations are faster than CPS
translating the entire code base (\cite[\S7]{caml-shift} and
references therein).  This is because CPS translating the entire
program allocates a great amount of intermediate closures, which OCaml
does not aggressively optimize. The direct-style implementation
thereby offers backwards compatible performance; only the code that
uses continuations pays the cost of creating and managing
continuations. The rest of the code behaves similar to vanilla OCaml.

Multicore OCaml does include support for multi-shot continuations, by allowing
the programmer to clone the continuation on-demand. Thus, the real
example |test1|
is implemented in multicore OCaml as,
\begin{code}
match f () with
| x -> x (* value clause *)
| effect Choose(x,y) k ->
    continue (Obj.clone_continuation k) x;
    continue k y
| effect Fail _ -> ()
\end{code}
In the above, we clone the continuation |k| using |Obj.clone_continuation|,
resume the continuation with |x| before resuming with |y|.

In Multicore OCaml, the program stack is linked list of stack
segments, where each segment is an object on the heap. Each segment
corresponds to a computation delimited by effect handlers. Thus, the
length of the linked list of stack segments is equal to the number of
effect handlers dynamically enclosing the current computation. Each
stack segment includes a slop space for the stack to grow. If the
stack overflows, we reallocate the stack segment in an object with
twice as much space as the original segment. The original stack
segment will eventually be garbage collected.

Since continuations are one-shot, capturing a continuation involves no
copying.  We need only to create a small object that points to a list
of stack segments that correspond to the continuation. Cloning a
continuation deep-copies the list of stack segments, and thereby
allows the same continuation to be resumed more than once. Multicore
OCaml's stack management is similar to the |Thread| module
implementation in the MLton Standard ML compiler in that both runtimes
manage stacks as dynamically resized heap objects. But they also
differ from each other since the continuations in MLton are
undelimited while they are delimited in Multicore OCaml. Clinger et
al.~\cite{Clinger1999-HOSC} describe various strategies for
implementing first-class undelimited continuations, which could be
adapted for delimited continuations. Multicore OCaml differs from all
these strategies in that the continuation is only copied if it is
explicitly demanded to be cloned. This decision makes the default case
of a continuation resumed exactly once fast.

\subsubsection{Delimcc in multicore OCaml}

We now discuss the Eff embedding in multicore OCaml. We achieve the embedding
by embedding the delimcc operators |new_prompt|, |push_prompt|, and |shift0| in
multicore OCaml. The embedding is given in Fig.~\ref{fig:delimcc_of_multicore}.
The |prompt| type is a record with two operations, one to take a
sub-continuation and the other to push a new prompt. We instantiate a new
prompt by declaring a new effect called |Prompt| in a local module. Thus, we
get a new |Prompt| effect instance for every invocation of
|new_prompt|. (The signature is written in a strange way as
|let new_prompt (type a) : unit -> a prompt| rather than the expected
|let new_prompt : 'a. unit -> 'a prompt|. The two notations are
equivalent, as far as the
user of |new_prompt| is concerned and describe the same polymorphic
type. However,
the former, by introducing a so-called ``locally abstract type'', lets
us use the type within |new_prompt|'s body, in the type annotation to
|effect Prompt|.) The |take|
operation wraps the given function |f| in the effect constructor and performs
it. The |push| operation evaluates |f| in a handler which handles the |Prompt|
effect. This handler applies the continuation to the given function |f|.

\begin{figure}[h!t]
\begin{code}
module type Delimcc = sig
  type 'a prompt

  val new_prompt   : unit -> 'a prompt
  val push_prompt  : 'a prompt -> (unit -> 'a) -> 'a
  val shift0           : 'a prompt -> (('b -> 'a) -> 'a) -> 'b
end

module Delimcc : Delimcc = struct
  type 'a prompt = {
          take  : 'b. (('b, 'a) continuation -> 'a) -> 'b;
          push : (unit -> 'a) -> 'a;
  }

  let new_prompt (type a) : unit -> a prompt = fun () ->
     let module M = struct effect Prompt : (('b,a) continuation -> a) -> 'b end in
     let take f  = perform (M.Prompt f) in
     let push th = match th () with
     | v -> v
     | effect (M.Prompt f) k -> f k
     in
     { take; push }

  let push_prompt  {push} = push

  let take_subcont {take} = take

  let push_subcont k v =
     let k' = Obj.clone_continuation k in
     continue k' v

  let shift0 p f =
     take_subcont p (fun sk -> f (fun c -> push_subcont sk c))
end
\end{code}
\caption{Embedding Delimcc in multicore OCaml}
\label{fig:delimcc_of_multicore}
\end{figure}

Now, the |push_prompt| and |take_subcont| operations are simply the
definitions of |push| and |take|, respectively. |push_subcont|
unconditionally clones the continuation and resumes it. Cloning is
necessary here since delimcc continuations are multi-shot. Finally,
|shift0| is implemented in terms of the operations to take and push
continuations, following its standard definition
\cite{dybvig-monadic,caml-shift} (see also \S\ref{s:sem-delimcc} for a
reminder). Since the handlers in Multicore OCaml are deep, the handler
installed at the corresponding |push_prompt| wraps the continuation
|sk|. If the continuations were \emph{shallow}, where the handler does
not wrap the continuation, the |shift0| encoding would be:
\begin{code}
let shift0 p f =
  take_subcont p (fun sk ->
	f (fun c -> push_prompt p (fun () -> push_subcont sk c)))
\end{code}
Thus, we have embedded in multicore OCaml a subset of Delimcc
operators used for our Eff embedding~-- and gained an
embedding of Eff in multicore OCaml.

\section{Eff in OCaml, Formally}
\label{s:formal}

In this section we formally state our translation from Eff to OCaml
and argue that it is meaning-preserving. First we recall the denotational
semantics of Eff. It is given in terms of OCaml values
rather than common denotational domains; \S\ref{s:what-is-denot}
discusses such style of denotations in more detail.
\S\ref{s:sem-delimcc} outlines the (novel) denotational
semantics of multi-prompt delimited control, in the style used previously in
\S\ref{s:sem-eff} for Eff. Finally, \S\ref{s:translation}
defines the translation, and argues that it preserves the
denotation of expressions.

\subsection{The Semantics of Eff}
\label{s:sem-eff}

The Eff paper \cite{eff-language} also introduced the language
formally, by specifying its denotational semantics. We recall it in this
section for ease of reference, making small notational adjustments for
consistency with the formalization of delimited control in the later
section.

\subsubsection{Core Eff}
\label{s:core-eff}

For ease of formalization and understanding, we simplify the
language to its bare minimum, Core Eff, presented in
Fig.~\ref{f:core-eff}.

\begin{figure}[htbp]
\begin{tabular}[Ct]{ll}
Variables & |x,y,z,u,f,k,r|\ldots \\
Constants & |c ::=| unit, integers, integer operations\\
Types &
\begin{code}
t ::= unit | int | t -> t | t +-> t | t => t
\end{code}
\\[2\jot]
Values &
\begin{code}
v ::= x | c | fnx:t. e | op v | h
\end{code}
\\
Handler &
\begin{code}
h ::= handler v (r -> e) ((x,k) -> e)
\end{code}
\\
Expressions (Computations) &
\begin{code}
e ::= val v | let x = e in e | v v | newp | with h handle e
\end{code}
\\
\end{tabular}
\caption{The Core Eff}
\label{f:core-eff}
\end{figure}

Whereas Eff, as a practical language, has a number of syntactic forms,
we limit Core Eff to the basic abstractions, applications and
let-expressions and use only the |with h handle e| notation for
effect handling. As in \cite{eff-language}, both components of an
application expression must be values.
Effects in Core Eff have only one 
operation (discussed in detail below) so it does not have to be named and
declared. Therefore, the simple |newp| expression suffices to
create effect instances, or values of effect types |t +-> t|. If |v'| is
an instance of an effect with, say, an integer argument, its
invocation is expressed as |op v' 1|, to be understood as the
application of the functional value |op v'| to the argument value |1|.
Besides the effect |t +-> t| and handler |t => t| types, Core Eff has only
unit, integer and function types. Other basic types, as well as products and
sums present in the full Eff are straightforward to add and their
treatment is standard. Therefore, we elide them. Values of handler types
are created by the form
|handler v (r -> e1) ((x,k) -> e2)|, where |v| must be an effect
instance. When the handled expression finishes normally, |e1| is evaluated with
the variable |r| bound to the expression's result.
If the handled expression invokes an effect, |e2| is evaluated with the
variable |x| bound to the effect argument and |k| bound to the continuation;
see \S\ref{s:core-eff-tagless-final} for the concrete example.
The |handler|
construct in Eff has a |finally| clause similar to the
|try ... finally| form found in many programming languages, to
post-process the result of the handler expression. This clause
is syntactic sugar and omitted in Core Eff.

Declaring several operations for an effect is certainly natural and
convenient. It turns out however that one can do without: no
expressiveness is lost if an effect has only one operation. Although
obvious in hindsight, this assertion seems surprising, even wrong.
Let's consider an Eff effect with three operations |o1|, |o2| and |o3| and let
|r| be its instance. In the following code (suggested by an anonymous
reviewer)
\begin{codeE}
handle e with
  val x     -> x
  r#o1 x k -> e1
  r#o2 x k -> e2
\end{codeE}
if |e| invokes |o3|, it is not handled by the shown handler and is
passed over (re-raised) to some outer handler.  Whenever expressions
|e1| or |e2| invoke any of the three operations, they, too, are to be
dealt with by that outer handler. Finally, when |e1| or |e2| invoke
the continuation |k| and, as the consequence, an operation |o1| or
|o2| is invoked, it will be dealt with again by |e1| (resp. |e2|). It
seems very difficult to locally, without global program rewriting,
emulate all that behavior using only a single-operation effect.

Yet such local emulation is possible~-- and, in hindsight, obvious.
An effect with multiple operations, for example,
\begin{codeE}
type exeff =
effect
  operation flip:      unit -> bool
  operation cow:     string -> string
  operation choose: (int * int) -> int
end
let r = new exeff
\end{codeE}
is equivalent to the effect with the single, `union' operation
\begin{codeE}
type uin   = InOp1 of unit | InOp2 of string | InOp3 of int * int
type uout = OutOp1 of bool | OutOp2 of string | OutOp3 of int
type eff1  = effect  operation op: uin -> uout  end

let r1 = new eff1
let r1flip x    = match r1#op (InOp1 x) with OutOp1 y -> y
let r1cow x    = match r1#op (InOp2 x) with OutOp2 y -> y
let r1choose x = match r1#op (InOp3 x) with OutOp3 y -> y
\end{codeE}
That is, |r#flip| is equivalent to |r1flip|, |r#choose| to
|r1choose|, etc., provided that the handlers are appropriately
adjusted. For example,
\begin{codeE}
handle e with
 | val x -> ev
 | r#flip x k   -> eflip
 | r#cow  x k -> ecow
\end{codeE}
is to be re-written as
\begin{codeE}
handle e with
 | val x -> ev
 | r1#op x k ->
   (match x with
      InOp1 x -> let k y = k (OutOp1 y) in eflip
    | InOp2 x -> let k y = k (OutOp2 y) in ecow
    | _        -> k (r1#op x))  (* default clause: re-raising *)
\end{codeE}
The accompanying code |many_one.eff| gives two complete examples,
including nested handlers. The shown re-writing of a multi-operation
effect into a single-operation one is local. The union data types can
be emulated with functions in Core Eff. The re-writing is also
cumbersome: one should take care to properly match the |InOp1| tag
with the |OutOp1| tag, etc. We should keep in mind however that Core
Eff is designed as an intermediate language and to simplify reasoning;
it is not meant for end-users.

The reliance on ordinary variant data types in our emulation gives the
impression of `lax typing' (excusable in an intermediate
language). It should be stressed however that
any algebraic signature can be properly represented as a data type
without any sloppiness, by means of generalized algebraic data types
(GADTs), created for that purpose \cite{xi-guarded}. (We did not use
GADTs here for the sake of simplicity.)

The single-operation encoding of multi-operation effects
should now become obvious. One sees the close analogy with
ordinary exceptions: multiple exceptions are usually implemented as a
single exception whose payload is an (extensible) union data type.
We also notice that extensible-effects in Haskell \cite{freer} are
based on the very same idea, implemented with no typing compromises.

We may also `split' a multiple-operation effect into multiple
single-operation effects. Taking the earlier |exeff| with |flip|,
|cow| and |choose| operations as an example, we define three new
single-operation effects:
\begin{codeE}
type eff_flip     = effect  operation op1:        unit -> bool   end
type eff_cow    = effect  operation op2:      string -> string end
type eff_choose = effect  operation op3: (int * int) -> int    end

let rflip     = new eff_flip
let rcow    = new eff_cow
let rchoose = new eff_choose
\end{codeE}
and replace |r#flip| with |rflip#op1|, |r#choose| with |rchoose#op3|,
etc. We still need the union data types |uin| and |uout| and
the unified effect |eff1|. In addition we define
\begin{codeE}
let flip_handler = handler
  | val x -> x
  | rflip#op1 x k  -> match r1#op (InOp1 x) with OutOp1 y -> k y
\end{codeE}
and similarly |cow_handler| and |choose_handler|. The old handlers are
re-written as in the previous method; in addition, we precompose the
|eff1| handlers with |flip_handler . cow_handler . choose_handler|.
The latter effectively converts three distinct effects into one single
|eff1|. This reification procedure also lets us emulate
multiple-effect Eff handlers such as
\begin{codeE}
handle e with
| rflip#op1 x k  -> ...
| rcow#op2 x k  -> ...
\end{codeE}
with only single-effect single-operation handlers.

All in all, in Core Eff an effect has only one operation, which hence
does not have to be named. There is no need for effect declarations
either. We do retain the facility to create, at run time, arbitrarily
many instances of the effect. In Core Eff, an effect instance alone
acts as the effect name.

Thus, Core Eff has unit, integer and arrow types, the type |t1 +-> t2|
of an effect operation that takes a |t1| value as an argument and produces
the result of the type |t2|, and the type |t1 => t2| of a handler acting on
computations of the type |t1| and producing computations of the type
|t2|.

\subsubsection{Core Eff in the Tagless-Final Form}
\label{s:core-eff-tagless-final}

The conventional presentation of syntax in Fig.~\ref{f:core-eff} can
be also given in a `machine-readable' form, as an OCaml
module signature, Fig.~\ref{f:eff-sig}.
\begin{figure}[h!t]
\begin{code}
module type Eff = sig
  type 'a repr                          (* type of values *)
  type 'a res                           (* type of computations *)

  type ('a,'b) eff                       (* effect instance type *)
  type ('a,'b) effh                      (* effect handler type *)

  (* values *)
  val int: int -> int repr
  val add: (int->int->int) repr
  val unit: unit repr

  val abs: ('a repr -> 'b res) -> ('a->'b) repr
  val op: ('a,'b) eff repr -> ('a ->'b) repr      (* effect invocation *)
  val handler: ('a,'b) eff repr ->                 (* effect instance *)
               ('c repr -> 'w res) ->               (* val handler *)
               ('a repr * ('b -> 'w) repr -> 'w res) -> (* operation handler *)
               ('c,'w) effh repr

  (* computations *)
  val vl: 'a repr -> 'a res                     (* all values are computations*)
  val let_: 'a res -> ('a repr -> 'b res) -> 'b res
  val ($$): ('a -> 'b) repr -> 'a repr -> 'b res
  val newp: unit -> ('a,'b) eff res             (* new effect instance *)
  val handle: ('c,'w) effh repr -> 'c res -> 'w res
end
\end{code}
\caption{The syntax and the static semantics of Core Eff, in the OCaml
  notation}
\label{f:eff-sig}
\end{figure}
The (abstract) OCaml type |'a repr| represents Core Eff type
|'a| of its values. In the same vein, |'a res| represents the type |'a|
of Eff computations. The paper \cite{eff-language} likewise
distinguishes the typing of values and computations, but in the form
of two different judgments\footnote{Since the signature \textsf{Eff}
  also represents the type system of Eff, in the natural deduction
  style, one may say that $\alpha$ \textsf{repr} and $\alpha$
  \textsf{res} represent a type judgment rather than a mere type.}.
A few concessions had to be made to OCaml syntax:
We write |(t1,t2) eff| for the effect type |t1 +-> t2| and
|(t1,t2) effh| for the type |t1 => t2| of handlers. We use |vl| in
OCaml rather than \textsf{val} since the latter is a reserved
identifier in OCaml. Likewise we spell Eff's \textsf{let} as |let_|,
the Eff application as the infix |$$|, and
give |newp| a dummy argument. We mark integer
literals explicitly: whereas |1:int| is an OCaml integer,
|(int 1):int repr| is Core Eff integer literal, which is the Eff
value of the Eff type |int|. We rely on higher-order abstract
syntax (HOAS)
\cite{Huet-HOAS,miller-manipulating,church-formulation},
using OCaml functions to represent Eff functions (hence
using OCaml variables for Eff variables).

The signature |Eff| encodes not just the syntax of Core Eff but
also its type system, in the natural-deduction style. For example,
the |val op| and |val handle| declarations in the |Eff| signature
straightforwardly represent
the following typing rules from \cite[\S3]{eff-language}, adjusted for
Core Eff and the natural deduction presentation:

\medskip
\begin{prooftree}
\vdash_v \ \textsf{v} : \textsf{t}_1\hookrightarrow\textsf{t}_2
\justifies
\vdash_v \ \textsf{op v} : \textsf{t}_1\to\textsf{t}_2
\end{prooftree}
~~~~~~
\begin{prooftree}
\vdash_v \ \textsf{h} : \textsf{t}_1\Rightarrow\textsf{t}_2
\qquad
\vdash_e \ \textsf{e} : \textsf{t}_1
\justifies
\vdash_e \ \textsf{with h handle e} : \textsf{t}_2
\end{prooftree}
\medskip

The type system has two sorts of judgments, for values
$\vdash_v\ \textsf{v}:\textsf{t}$ and for computations
$\vdash_e\ \textsf{e}:\textsf{t}$~-- which we distinguish by giving the
type |t repr| to the encoding of Eff values and |t res| to the
encoding of computations. The rules express the intent that effect
operation invocations act as functions and that a handler acts as an
expression transformer.

The benefit of expressing the syntax and the type system of a language in
the form of an |Eff|-like signature~-- in the so-called
\emph{tagless-final style} \cite{carette-finally,tagless-final-oxford}~--
and the reason to tolerate concessions to OCaml syntax is the ability
to write core Eff code and have it automatically typed-checked (and
even getting the types inferred) by the OCaml type checker.

As an illustration, we define a Reader-like |int +-> int| effect that
increments its
argument by the value passed in the environment. The |ans| expression invokes
the operation twice on the integer |1|, eventually supplying |10| as
the environment; the expected result is |21|. In Core Eff (or, to be precise,
the subset of Eff equivalent to Core Eff), the example looks as follows.
The responses of the Eff interpreter are shown in the comments.

\begin{codeE}
type reader =
effect
  operation op : int -> int
end

let readerh p = handler
  | val v    -> (fun s -> v)
  | p#op x k -> (fun s -> let z = s + x in k z s)
(* val readerh : reader ->  ('a =>  (int ->  'a)) = <fun> *)

let ans =
   let p = new reader in
   (with readerh p handle
      let x = p#op 1 in
      let y = p#op x in
      y
   ) 10 (* the value passed in the environment *)
(* val ans : int = 21 *)
\end{codeE}

The tagless-final encoding of the same example is:
\begin{code}
module Ex1(E:Eff) = struct
  open E
  (* A macro to apply a computation: mere ($$) applies a value *)
  let ($$$) e x = let_ e (fun z -> z $$ x)

  let readerh p = handler p 
   (fun v     -> vl @@ abs (fun s -> vl v))
   (fun (x,k) -> vl @@ abs (fun s ->
                   let_ ((add $$ s) $$$ x) (fun z -> (k $$ z) $$$ s )))

  let ans =
    let_ (newp ()) @@ fun p ->
    let_ (handle (readerh p) @@
           let_ (op p $$ (int 1)) (fun x ->
           let_ (op p $$ x)       (fun y ->
           vl y))) (fun hr ->
    hr $$ int 10)
end
\end{code}
The OCaml type-checker verifies the code is type-correct and infers
for |ans| the type |int E.res|, meaning |ans| is a computation
returning an |int|. For |readerh|, the type
|(int, int) eff repr| |-> ('a, int -> 'a) effh repr| is inferred, which
corresponds exactly to the inferred type of Eff's |readerh|.

\subsubsection{`Interpreter-based' Denotational Semantics of Core Eff}
\label{s:core-eff-denot}

There is another significant benefit of the tagless-final style. The
signature |Eff| looks like a specification of a denotational semantics
for the language. Indeed, |repr| and |res| look like semantic
domains~-- corresponding to the domains |V| and |R| from
\cite[\S4]{eff-language}, but indexed by types. Then |int|, |abs|, |op|,
|handler| and the other members of the |Eff| signature are the semantic
functions, which tell the meaning of the corresponding Eff value or
expression from the meaning of its components. The
compositionality is built into the tagless-final approach.

The signature |Eff| is only the specification of semantic functions. To
define the denotational semantics of Core Eff we need to give the
implementation. It is shown in Fig.~\ref{f:eff-sem}.
\begin{figure}[tp]
\begin{code}
module REff = struct
  type 'a repr  =
    | B : 'a -> 'a repr                          (* OCaml values *)
    | F : ('a repr->'b res) -> ('a->'b) repr  (* Functions V->R,
                                                         i_arr in the Eff paper *)
  and _ res   =                                 (* Results *)
    | V: 'w repr -> 'w res                      (* Normal termination result *)
    | E: {inst: int; arg:'a repr; k:'b repr->'w res} -> 'w res

  let rec lift : ('a repr -> 'b res) -> 'a res -> 'b res = fun f -> function
    | V v -> f v
    | E ({k;_} as oper) -> E {oper with k = fun x -> lift f (k x)}

  type ('a,'b) eff = int
  type ('a,'b) effh = (unit -> 'a) -> 'b

  (* values *)
  let int (x:int) = B x
  let add : (int->int->int) repr =
    F (function B x -> V (F (function B y -> V (B (x+y)))))
  let unit = B ()

  let abs f = F f
  let ($$): ('a -> 'b) repr -> 'a repr -> 'b res =
    function F f -> fun x -> f x

  let op: ('a,'b) eff repr -> ('a -> 'b) repr = function B p ->
    abs (fun v -> E {inst=p; arg=v; k = fun x -> V x})

  let handler: ('a,'b) eff repr ->                    (* effect instance *)
               ('c repr -> 'w res) ->                  (* val handler *)
               ('a repr * ('b -> 'w) repr -> 'w res) -> (* operation handler *)
               ('c,'w) effh repr =
     fun (B p) valh oph ->
       let rec h = function
         | V v            -> valh v
         | E {inst;arg;k} when inst = p ->
             (* if inst = p then arg and k have specific types recovered below *)
             let (arg:'a repr)        = Obj.magic arg in 
             let (k:'b repr -> 'c res) = Obj.magic k in
             (* Since the handlers are deep, we compose k with h *)
             oph (arg, abs (fun b -> h (k b)))
         (* Relay to an outer handler *)
         | E ({k:_} as oper) -> E {oper with k = fun b -> h (k b)}
       in abs (fun th -> h (th $$ unit))

  let vl v = V v                     (* all values are computations *)
  let let_: 'a res -> ('a repr -> 'b res) -> 'b res = fun e f -> lift f e

  let newp: unit -> ('a,'b) eff res =
    let c = ref 0 in
    fun () -> incr c; V (B !c)

  let handle: ('c,'w) effh repr -> 'c res -> 'w res =
    fun h e -> h $$ abs (fun (_:unit repr) -> e)
end
\end{code}
\caption{The denotational semantics of Core Eff}
\label{f:eff-sem}
\end{figure}
The module |R| implementing |Eff| is essentially the denotational
semantics of Eff given in \cite[\S4]{eff-language}, but written in a
different language: OCaml rather than the standard mathematical
notation.  It is undeniable that the conventional mathematical
notation is concise~-- although the conciseness comes in part from
massive overloading and even sloppiness, omitting details like various
inclusions and retractions. The OCaml notation is precise. Moreover,
the OCaml type-checker will guard against typos and silly mistakes. Since
we index the domains by type, there are quite a few simple
correctness properties that can be ensured effectively and simply.
For example, forgetting to compose the continuation with the handler
|h| in |handler| leads to a type error. We discuss this style of
denotation in more detail in \S\ref{s:what-is-denot}.

The denotations of Core Eff are expressed in terms of two semantic
domains, of values and results. In \cite{eff-language}, the domains
are called |V| and |R| respectively. We call them |'a repr| and
|'a res|, and index by types. The type-indexing lets us avoid many of
the explicit inclusions and retractions defined in
\cite[\S4]{eff-language}.  In our |R|
implementation, domains are defined concretely, as OCaml values, viz.
mutually recursive data types |repr| and |res|. Of all the retracts of
\cite{eff-language} we only need two non-trivial ones. The first is
$\rho_\rightarrow$ in
\cite{eff-language} (with the corresponding inclusion $\iota_\rightarrow$),
which embeds the
functions |'a repr -> 'b res| into |('a->'b) repr|. This embedding is
notated as |F| (the inclusion is applying the |F| constructor and
the retraction is pattern-matching on it). The second retract
deals with the embedding of
|'a res -> 'b res|: such functions are isomorphic to
|(unit->'a) repr -> 'b res|, which are then embedded into
|((unit->'a)->'b) repr| as described earlier. The domain |repr| does not need
the bottom element since values are vacuously terminating, and
our denotational semantics is
typed, Church-style: we give meaning only to well-formed and
well-typed expressions.

We define the domain |'a res| to be nothing bigger than its required
retract, the sum expressing the idea that a terminating computation is
either a value, |V|, or an effect operation. The latter is a tuple
that collects all data about the operation: the instance, the
argument, and the continuation. The lifting of |f:'a repr -> 'b res|
to the |'a res| domain, written as $f^\dagger$ in \cite{eff-language},
is notated as |lift f| in our presentation.  The implementation of
|int|, |abs|, |op| and the rest of the members of |Eff| is the
straightforward transcription of the definitions from
\cite{eff-language}. (We use the higher-order abstract syntax and
hence do not need the explicit `environment' $\eta$.)  The appearance
of |Obj.magic| comes from the fact that Core Eff (just like the
full Eff) does not carry the effect type in the type of a
computation. Therefore, the argument and result types of an effect are
existentialized. One may hence view |Obj.magic| as an implicit
retraction into the appropriate |'a repr| domain. The use of
|Obj.magic| is safe, thanks to the property that each effect instance (denoted
by an integer) is unique; that is, the instances of differently-typed
effects have distinct values.

Having recalled the semantics of Eff, we now turn to the delimited
control, and then, in \S\ref{s:translation}, to the translation from
Core Eff to Core OCaml with delimited control.

\subsubsection{Digression: What is Denotational Semantics?}
\label{s:what-is-denot}

The semantics just presented in \S\ref{s:core-eff-denot} may raise
eyebrows: one commonly thinks of denotational semantics as giving
interpretations through mathematical
objects rather than OCaml code. It is worth therefore, to take a moment
to reflect on what exactly denotational semantics is.

One of the first definitions of denotational semantics (along with
many other firsts) is given by Landin: \cite[\S8]{landin-700}
\begin{quotation}
``The commonplace expressions of arithmetic and algebra have a certain
  simplicity that most communications to computers lack. In
  particular, (a) each expression has a nesting subexpression
  structure, (b) each subexpression denotes something (usually a
  number, truth value or numerical function), (c) the thing an
  expression denotes, i.e., its `value', depends only on the values of
  its subexpressions, not on other properties of them.''
\end{quotation}
As an illustration, Landin then describes the denotations of string
expressions in terms of (natural language) strings such as `wine' or
even equivalence classes of ISWIM-like expressions.

In the reference text \cite[\S3.1]{Mosses-denot}, Mosses essentially
repeats Landin's definition, adding: ``It should be noted
that the semantic analyst is free to \emph{choose} the denotations of
phrases~-- subject to compositionality''. He notes that letting
phrases denote themselves is technically compositional and hence may
be accepted as a denotational semantics~-- which however has
``(extremely) poor abstractness''. Still, he continues, there are two
cases where it is desirable to use phrases as denotations, e.g., for
identifiers.

Thus from the very beginning there has been precedent of using
something other than abstract mathematical sets or domains as
denotations. Even syntactic objects may be used for semantics,
provided the compositionality principle is satisfied. In this paper,
we take as semantic objects OCaml values, equipped with
\emph{extensional} equality. In case of functions, checking the
equality involves reasoning if two OCaml functions, when applied to
the same arguments, return the extensionally equal results. To be more
precise, we check how the OCaml (byte-code) interpreter evaluates the
applications of these functions to the same arguments. The behavior of
the byte-code interpreter is well-defined; the compilation of the
fragment of OCaml we are using is also well-understood (including
|Obj.magic|, which operationally is the identity). We give an example
of such reasoning in \S\ref{s:sem-delimcc-adequacy}.

Using an interpreter to define a language has long precedent, starting
from Reynolds' \cite{reynolds-definitional}. Such an approach was also
mentioned by Schmidt in the survey \cite{Schmidt-denot}:
\begin{quotation}
	``A pragmatist might view an operational or denotational
semantics as merely an `interpreter' for a programming language. Thus,
to define a semantics for a general-purpose programming language, one
writes an interpreter that manipulates data structures like symbol
tables (environments) and storage vectors (stores).
For example, a denotational semantics for an imperative
language might use an environment, |e|, and a store, |s|, along with an
environment lookup operation, |find|, and a storage update operation,
|update|. Since data structures like symbol tables and storage vectors
are explicit, a language's subtleties are stated clearly and its flaws
are exposed as awkward codings in the semantics.''
\end{quotation}

\subsection{Denotation of Delimited Control}
\label{s:sem-delimcc}

This section describes the target language of the Eff embedding,
which is OCaml with the delimcc library. As we did with Eff, we reduce
the language to the bare minimum, to be called Core delimcc. The
syntax and the static semantics (that is, the type system) is
presented in Fig.~\ref{f:delimcc-sig}. From now on, we will be
using the OCaml rather than the mathematical notation~-- as was
first presented in \S\ref{s:sem-eff}.

The Core delimcc is, in many parts, just like Core Eff,
Fig.~\ref{f:eff-sig}, and is likewise described by the OCaml
signature.  The Core delimcc is a bigger language: we need enough
features to be able to write |handle_it| from
\S\ref{s:eff-delimcc}. Therefore, besides ordinary function
definitions, Core delimcc has recursive functions |absrec|. Recursive
functions can also be defined in the full Eff; we did not need them
for the Core Eff subset. The (user-defined) |'e result| data type of
\S\ref{s:eff-delimcc} is built into Core delimcc as
|free|, which is a sum whose second summand is a tuple. The data type
is represented by the constructors |ret| and |act| for the summands,
and the deconstructor (eliminator) |with_free|\footnote{\relax
Therefore, \textsf{free} could have been left in the signature as an
abstract type. We gave the full data type declaration instead
because it seems instructive, making it easier to understand the
types of the constructors and the deconstructor.}. For simplicity we
chose the |'e result_v1| version of the result data type, with the universal
type (rather than the more complicated out-of-band carrying of normal
computational results). Therefore, Core delimcc has the universal
type with the corresponding
injection |i_univ| and projection |p_univ|. The delimcc-specific
part \cite{caml-shift} is the type of control delimiters, so-called
prompts, the operations to create a fresh prompt |newpr|, set the
prompt |pushpr| and to capture the continuation up to the dynamically
closest |pushpr|, the operation ``shift-0'' |sh0|. (Other than
this delimcc-specific part, the rest of the |Delimcc| signature is, strictly
speaking, mere for the sake of the Eff embedding. However, as
we saw already in \S\ref{s:eff-delimcc}, the universal type and
something like the |free| data type often come up when using delimcc.)

\begin{figure}[h!!t]
\begin{code}
module type Delimcc = sig
  type 'a repr
  type 'a res

  (* values *)
  val int: int -> int repr
  val add: (int->int->int) repr
  val unit: unit repr

  type univ                             (* the universal type *)
  val i_univ: 'a repr -> univ repr
  val p_univ: univ repr -> 'a res

  val abs: ('a repr -> 'b res) -> ('a->'b) repr
  val absrec: (('a->'b) repr -> 'a repr -> 'b res) -> ('a->'b) repr

  type ('a,'b) free =
    | Ret of univ repr
    | Act of 'a repr * ('b -> ('a,'b) free) repr
  val ret: univ repr -> ('a,'b) free repr
  val act: 'a repr -> ('b -> ('a,'b) free) repr -> ('a,'b) free repr
  val with_free: ('a,'b) free repr ->
                 (univ repr -> 'w res) ->
                 ('a repr -> ('b -> ('a,'b) free) repr -> 'w res) ->
                 'w res

  (* computations *)
  val vl: 'a repr -> 'a res                     (* all values are computations*)
  val let_: 'a res -> ('a repr -> 'b res) -> 'b res
  val ($$): ('a -> 'b) repr -> 'a repr -> 'b res

  (* The delimcc part: prompt and shift *)
  type 'a prompt
  val newpr: unit -> 'a prompt res
  val pushpr: 'a prompt repr -> 'a res -> 'a res
  val sh0:    'a prompt repr -> (('b -> 'a) repr -> 'a res) -> 'b res
end
\end{code}
\caption{The syntax and the type system of Core delimcc}
\label{f:delimcc-sig}
\end{figure}

Like Core Eff \S\ref{s:sem-eff}, Core delimcc distinguishes the type
of values from the type of computations. In this we squarely follow
the lead of Bauer and Pretnar \cite{eff-language}: whereas the
user-visible Eff, like the real OCaml, does not distinguish effectful
computations from values in its types, the formal presentation of
Eff in \cite{eff-language} does, in syntax, in type system, and
dynamic semantics. One may think of Core delimcc as the A-normal form of
OCaml delimcc. To better see the correspondence, we take one, rather
advanced example of the delimcc OCaml code (from the delimcc test suite),
featuring several prompts and the repeated invocations of captured
continuations
\begin{code}
let p1 = new_prompt () in
let p2 = new_prompt () in
let p3 = new_prompt () in
let pushtwice sk =
  sk (fun () ->
    sk (fun () ->
      shift0 p2 (fun sk2 -> sk2 (fun () -> sk2 (fun () -> 3))) ())) in
push_prompt p1 (fun () ->
  push_prompt p2 (fun () ->
    push_prompt p3 (fun () -> shift0 p1 pushtwice ()) + 10) + 1) + 100
\end{code}
We re-write the example in Core delimcc as follows
\begin{code}
module ExD(D:Delimcc) = struct
 open D

  (* A macro to apply to computation: ($$) applies to value *)
 let ($$$) e x = let_ e (fun z -> z $$ x)

 let (++) e v = let_ e (fun ev -> let_ (add $$ ev) (fun fv -> fv $$ v))

 let ans =
  let_ (newpr ()) @@ fun p1 ->
  let_ (newpr ()) @@ fun p2 ->
  let_ (newpr ()) @@ fun p3 ->
  let pushtwice sk = (* OCaml let: macro *)
    sk $$ abs (fun (_:unit repr) ->
      sk $$ abs (fun (_:unit repr) ->
	sh0 p2 (fun sk2 -> sk2 $$ abs (fun (_:unit repr) ->
           sk2 $$ abs (fun (_:unit repr) -> vl (int 3)))) $$$ unit)) in
  pushpr p1 (
    pushpr p2 (
      pushpr p3 (sh0 p1 pushtwice $$$ unit) ++ int 10) ++ int 1) ++ int 100
end
\end{code}
After defining several `macros', the rewriting is systematic and
straightforward. The real OCaml delimcc relates to Core delimcc quite
like Eff relates to Core Eff as was illustrated in
\S\ref{s:core-eff-tagless-final}.

The semantics of delimited control is typically presented in the
small-step reduction style (see \cite{dybvig-monadic,caml-shift}):
\begin{code}
pushpr p (vl x)                         ### vl x
pushpr p (Cp[sh0 p (fun k -> e)]) ### let k = abs (fun x -> pushpr p Cp[x]) in e
\end{code}
where |Cp[]| is the evaluation context with no
sub-context |pushpr p []|. In contrast, we treat Core delimcc
denotationally, giving it semantics inspired by the ``bubble-up''
approach of \cite{felleisen-reasoning,parigot-lambda-mu}. We establish
the correspondence in \S\ref{s:sem-delimcc-adequacy}.

Our (interpreter-based) denotational semantics of Core delimcc,
Fig.~\ref{f:delimcc-sem}, is (intentionally)
quite similar to that for Core Eff, in Fig.~\ref{f:eff-sem}. It is
given in terms of domains |'a repr| of value denotations and
|'a res| of expression denotations. The value denotations are the same
as in Core Eff. A terminating expression is either a
value |V|,
or a ``bubble'' |E| created by |sh0|. The bubble merely packs the data
from the |sh0| that created it (the prompt value plus the body of the
|sh0| operator), along with the continuation |k| that represents the
context of that |sh0|. All in all, the bubble represents the
decomposition of an expression as the |sh0| operation embedded into an
evaluation context.

\begin{figure}[tp]
\begin{code}
module RDelimcc = struct
  type 'a repr  =
    | B : 'a -> 'a repr
    | F : ('a repr->'b res) -> ('a->'b) repr
  and _ res   =
    | V: 'a repr -> 'a res
    | E: {prompt: int; body:('c ->'w) repr->'w res; k: 'c repr -> 'a res} -> 'a res

  let rec lift : ('a repr -> 'b res) -> 'a res -> 'b res = fun f -> function
    | V v -> f v
    | E ({k;_} as oper) -> E {oper with k = fun c -> lift f (k c)}

  (* values *)
  let int (x:int) = B x
  let add : (int->int->int) repr = F (function B x -> V (F (function B y -> V (B (x+y)))))
  let unit = B ()

  type univ = Obj.t                            (* the universal type *)
  let i_univ: 'a repr -> univ repr = fun x -> B (Obj.repr x)
  let p_univ: univ repr -> 'a res  = function B x -> V (Obj.obj x)

  let abs f = F f
  let absrec: (('a->'b) repr -> 'a repr -> 'b res) -> ('a->'b) repr = fun f ->
    abs (fun x -> let rec h y = f (abs h) y in h x)

  let vl v = V v                     (* all values are computations *)
  let let_: 'a res -> ('a repr -> 'b res) -> 'b res = fun e f -> lift f e
  let ($$): ('a -> 'b) repr -> 'a repr -> 'b res = function F f -> fun x -> f x

  type ('a,'b) free =
    | Ret of univ repr
    | Act of 'a repr * ('b -> ('a,'b) free) repr
  let ret: univ repr -> ('a,'b) free repr = fun x -> B (Ret x)
  let act: 'a repr -> ('b -> ('a,'b) free) repr -> ('a,'b) free repr = fun v k -> B (Act (v,k))
  let with_free: ('a,'b) free repr -> (univ repr -> 'w res) ->
                 ('a repr -> ('b -> ('a,'b) free) repr -> 'w res) -> 'w res =
   fun (B free) reth acth -> match free with
   | Ret x -> reth x
   | Act (a,k) -> acth a k

  type 'a prompt = int
  let newpr: unit -> 'a prompt res =
    let c = ref 0 in
    fun () -> incr c; V (B !c)

  let sh0: 'a prompt repr -> (('b -> 'a) repr -> 'a res) -> 'b res =
    fun (B prompt) body -> E {prompt;body;k=vl}

  let rec pushpr: 'a prompt repr -> 'a res -> 'a res = fun (B p) -> function
    | V x -> V x
    | E{prompt; body;k} when prompt = p ->
        let (body:_->_) = Obj.magic body in
        body (abs (fun c -> pushpr (B p) (k c)))
         (* Relay to an outer handler *)
    | E ({k;_} as oper) -> E {oper with k = fun c -> pushpr (B p) (k c)}
end
\end{code}
\caption{The denotational semantics of Core delimcc}
\label{f:delimcc-sem}
\end{figure}

The only non-standard parts of the semantics are the denotations of
|sh0| and |pushpr|. As was already said, |sh0| creates the bubble, by
packing its arguments along with the identity continuation
representing the empty context. The function |lift| (essentially
|let_|)~-- which represents a let-bound expression in the context of the
let-body~-- grows the bubble by adding to it the let-body context.
The operation |pushpr p| ``pricks'' the
bubble (but only if the prompt value |p| matches the prompt value
packed inside the bubble, that is, the prompt value of the |sh0| that
created the bubble). When the bubble is pricked, the |sh0| body hidden
inside is released and is applied to the continuation accumulated
within the bubble~-- enclosed in |pushpr p| as behooves the shift
operation. Again, |Obj.magic| comes from the fact that we do not
carry the answer type in the type of a computation. Therefore,
the answer type |'w| is existentialized in the bubble. When the bubble
is pricked however, we are sure that the answer-type is actually the
type of the |pushpr| computation. The coercion operation is hence
safe. The |RDelimcc| implementation of the |Delimcc| signature lets us
run the example |ExD|, which gives |135| (the same result as the real
OCaml delimcc).

\subsubsection{Adequacy of the Core delimcc Semantics}
\label{s:sem-delimcc-adequacy}

As an illustration of the just defined interpreter-based denotational
semantics, and a quick check of its adequacy, we demonstrate that the
semantics models the key feature of the |shift0| control operator.

The behavior of |shift0| and its companion |push_prompt| is commonly
defined by the following re-writing (\cite{dybvig-monadic}, among
others)
\begin{code}
pushpr p (vl x)                         ### vl x
pushpr p (Cp[sh0 p (fun k -> e)]) ### let k = abs (fun x -> pushpr p Cp[x]) in e
\end{code}
mentioned earlier. Here |Cp[]| is the evaluation context with no
sub-context |pushpr p []|. We now show that these re-writing rules
preserve the denotation of expressions. In other words, the
left-hand-side and the right-hand-side of these (oriented) equations
have the same denotations. This is clearly the case for the first
re-writing rule. As far as the second rule is concerned, we take one
characteristic case, for one particular context |Cp[]|, namely,
|let opv = [] in let argv = arg in opv argv|, where |arg| is an
expression. The other cases are similar.

We shall thus show that the following two expressions have the same
denotations
\begin{code}
let el = pushpr p (let_ (sh0 p (fun k -> e)) (fun opv ->
                   let_ arg                  (fun argv ->
                   opv $$ argv)))
let er =
  let k = abs (fun x ->
               pushpr p (
                let_ (vl x) (fun opv ->
                let_ arg    (fun argv ->
                opv $$ argv))))
  in e
\end{code}
We will write |EV[e]| for the denotation of the Core delimcc
expression |e|, and, abusing the notation, |EV[v]| for the denotation
of the value |v|.
(Recall, for an expression |e| of the type |t|, 
|EV[e]| is an OCaml value of the type |t res|). Thus we demonstrate that
|EV[el] = EV[er]| for all expressions |e| and |arg| and the
value |p| of appropriate types.

Using the |RDelimcc| semantics, we build up the following denotations:
\begin{code}
EV[p] = B p'  /+where+/ p' /+is an integer+/
EV[sh0 p (fun k -> e)]
 = E{prompt=p'; body=fun k->EV[e]; k=fun v -> V v}
EV[let_ (sh0 p (fun k -> e)) (fun opv -> let_ arg (fun argv -> opv $$ argv))]
 {/+definition of+/ let_}
 = lift ctxf EV[(sh0 p (fun k -> e))]
 {/+definition of+/ lift}
 = E{prompt=p'; body=fun k->EV[e]; k=fun c -> lift ctxf (V c)}
 where
  ctxf = fun opv -> EV[let_ arg (fun argv -> opv $$ argv)]

EV[el]
 {/+definition of+/ pushpr; /+case of the matching prompt+/}
 = (fun k->EV[e])
     (abs (fun c -> EV[pushpr] (B p') (lift ctxf  (V c))))
 {/+definition of+/ let_}
 = (fun k->EV[e])
     (abs (fun c -> EV[pushpr] (B p') (EV[let_] (V c) ctxf)))
 = (fun k->EV[e])
     (abs (fun c -> EV[pushpr] (B p')
                      EV[let_ (vl c) (fun opv -> let_ arg (fun argv -> opv $$ argv))]))
 = EV[er]
\end{code}
We used the facts that, for example, |lift f e| can be substituted
with |EV[let_] e f| in all contexts: the left-hand-side of a
non-effectful let-definition is inter-substitutable with the right-hand-side,
preserving extensional equality.

One may also want to check the satisfaction of
the axioms \cite{Kameyama-axiom}; we leave it for future work.

\subsection{Translation from Eff to Delimited Control, and its
  Correctness}
\label{s:translation}

Having formalized the semantics of Core Eff and Core delimcc,
we are now in a position to formally state the
translation and argue about its correctness.

The tagless-final style used for the denotational semantics makes it
straightforward to express a compositional translation. Indeed, a
language is specified as an OCaml signature that collects the
declarations of syntactic forms of the language. The
interpretation~-- semantics~-- is an implementation of the
signature. A given signature may have several implementations. For
example, the signature |Eff| (Fig.~\ref{f:eff-sig})
of Core Eff had the |REff| implementation (Fig.~\ref{f:eff-sem}).
Fig.~\ref{f:translation} shows another
implementation, in terms of Core delimcc: it maps the types
and each of the primitive expression forms of Core Eff to the types
resp. expressions of Core delimcc. The mapping homomorphically
extends to composite Core Eff expressions; such an extension is
inherent in the tagless-final approach. The mapping should not depend
on any concrete implementation of delimcc: therefore, it is
formulated only in terms of the abstract types and methods defined in
the |Delimcc| signature, Fig.~\ref{f:delimcc-sig}. In OCaml terms, the
translation is represented as a functor, |Delimcc -> Eff|.

\begin{figure}[tp]
\begin{code}
module Translation(D:Delimcc) = struct
  type 'a repr  = 'a D.repr
  type 'a res   = 'a D.res

  type ('a,'b) eff  = ('a,'b) D.free D.prompt
  type ('a,'b) effh = ((unit -> 'a) -> 'b)

  (* values *)
  let int  = D.int
  let add  = D.add
  let unit = D.unit

  let abs = D.abs

  let op: ('a,'b) eff repr -> ('a -> 'b) repr = fun p ->
    D.(abs (fun v -> sh0 p (fun k -> vl @@ act v k)))

  let compose: ('b->'c) repr -> ('a->'b) repr -> ('a->'c) repr = fun fbc fab ->
    D.(abs (fun a -> let_ (fab $$ a) (fun b -> fbc $$ b)))

  let handler: ('a,'b) eff repr ->                    (* effect instance *)
               ('c repr -> 'w res) ->                 (* val handler *)
               ('a repr * ('b -> 'w) repr -> 'w res) -> (* operation handler *)
               ('c,'w) effh repr =
     fun p valh oph ->
       let h = D.(absrec @@ fun h freer -> 
         with_free freer 
           (fun r   -> let_ (p_univ r) (fun r -> valh r))
             (* Since the handlers are deep, we compose k with h *)
           (fun v k -> oph (v,compose h k)))
      in 
       D.(abs (fun th -> 
         let_ (pushpr p (let_ (th $$ unit) (fun r -> vl (ret (i_univ r)))))
         (fun freer -> h $$ freer)))

  let vl   = D.vl
  let let_ = D.let_
  let ($$) = D.($$)

  let newp: unit -> ('a,'b) eff res = D.newpr

  let handle: ('c,'w) effh repr -> 'c res -> 'w res =
    fun h e -> h $$ abs (fun (_:unit repr) -> e)
end
\end{code}
\caption{Translation from Core Eff to Core delimcc}
\label{f:translation}
\end{figure}

The translation is rather straightforward: |'a repr| and |'a res|
domains of Eff map to the corresponding domains of delimcc. An Eff effect
instance maps to a delimcc prompt. Most of Core Eff expression
forms (|int|, |add|, |abs|, |let_|, etc) map to the corresponding Core
delimcc forms. Only |op| and |handler| of Core Eff have
non-trivial implementation in terms of delimcc: |op| is just
|sh0| that creates a bubble with the data about the effect
operation. The effect handler interprets those data. Since effect
handlers in Eff are deep (that is, after an effect is handled and the
expression is resumed, the handler is implicitly re-applied), they
correspond to recursive functions in Core delimcc.  Again, the
appearance of the universal type in |handler| comes from the fact that
we do not carry the effect type in the type of a computation.
In \S\ref{s:eff-delimcc} we emulated the universal type in terms of
reference cells.

The |Translation| functor defines, in OCaml notation, the translation
of Core Eff types and expressions, which we can notate
|TRB t TRE| and |TRB e TRE|. The facts that the translation deals with typed
expressions and the |Translation| functor is accepted by the
OCaml type-checker immediately lead to:

\begin{proposition}[Type Preservation]
If |e| is a Core Eff expression of the type |t| (whose free variables,
if any, have the types |x1:t1,...|), then |TRB e TRE| has the type
|TRB t TRE| (assuming free variables of the types |x1:TRB t1 TRE,...|).
\end{proposition}
The proof immediately follows from the typing of the |Translation| functor.

We thus have two implementations of Core Eff: the original |REff|
(Fig.~\ref{f:eff-sem}) and the result of the translation
|Translation(RDelimcc)|. Before we can state the main theorem that
these two implementations are ``the same'' and hence the translation
is meaning-preserving, we have to verify that the semantic domains of
the two denotational semantics are comparable. The |REff|
implementation has |'a repr| and |'a res| domains defined in
Fig.~\ref{f:eff-sem} whereas the translated one uses |'a repr| and
|'a res| from Fig.~\ref{f:delimcc-sem}. While the two |'a repr| have the
same structure, |'a res| differ slightly. Both are sums, with the
identical |V| component, and the |E| component being a triple:
|{inst: int; arg:'c repr; k:'b repr->'a res}| vs.
|{prompt: int; body:('b ->'c) repr->'c res; k: 'b repr -> 'a res}|.
Although the first and the third components of the triple are
compatible, the middle is not. A moment of thought shows that the only
delimcc bubbles in the |Translation(RDelimcc)|
implementation are those that come from |op|, in which case the
body of the bubble is |fun k -> vl @@ act v k|, or, unfolding the
definitions, |fun k -> V (Act (v,k))|, with |v| being the argument
|arg| of the effect operation. Hence the triple
|{inst;arg;k}| can be turned to
|{prompt=inst;body = (fun k -> V (Act(arg,k)));k}| (and easily
retracted back). In the end, although |'a RDelimcc.res| domain is `bigger',
to the extent it is used in the |Translation(RDelimcc)|, it is
isomorphic to |'a REff.res|. This isomorphism is implicitly used in
the main theorem:

\begin{proposition}[Meaning Preservation]
A Core Eff value or expression has the same meaning (that is,
interpreted as extensionally the same OCaml value) under |REff| and
|Translation(RDelimcc)| semantics.
\end{proposition}
The proof has to verify that types correspond to the same domains in
both interpretations and that primitive forms of Core Eff have the
same interpretations. We have already discussed the |'a repr| and |'a res|
domains in both semantics. Clearly |('a,'b) eff| type has the same
interpretation (integer in both semantics), and so does |('a,'b) effh|.
Most of Core Eff forms have obviously the same interpretation in
both semantics. The only non-trivial argument concerns |op| and
|handler|. The expression |op p| denotes the function
|fun v -> E{inst=p;arg=v;k=fun x->V x}| under the |REff| semantics and
the function
|fun v -> E{prompt=p;body=(fun k -> V (Act(v,k)));k=fun x->V x}|
under the translation semantics. As we argued earlier, the denotations
are the same (keeping our isomorhism in mind).

The |handler p valh oph| in both interpretations is a function from |'c res|
to |'w res|. To see that it is the same function, we consider three cases.
First, if the argument is of the form |V x|, both interpretations
converge on |valh x|. If the argument is of the form
|E {inst;arg;k}| (in the |REff| interpretation) with |inst=p|,
the first interpretation gives
|oph (arg, handler p valh oph . k)|. In the translation interpretation,
the corresponding handled expression has the denotation
|E {prompt;body;k}|, with |prompt| being equal to |p| and
|body| being |fun k -> V (Act (arg,k))|.
Then |pushpr p (E {prompt;body;k})| amounts to
|body (pushpr p . k)|, which is |V (Act (arg,pushpr p . k))|. It is
then handed over to the recursive function |h| in
Fig.~\ref{f:translation}, which returns |oph (arg, h . pushpr p . k)|.
The latter matches the |REff| denotation.
The remaining case is of the handled expression being
|E {inst;arg;k}| (in the |REff| interpretation) with |inst| different
from the handler's |p|. The |REff| interpretation gives
|E {inst;arg;handler p valh oph . k}|. It is easy to see that the
translation interpretation gives the same.

\section{Higher-Order Effects}
\label{s:higher}

The running example from \S\ref{s:eff-intro} used the single instance
|r| of the |nondet| effect, created at the top level~-- essentially,
`statically'. Eff also supports creating effect instances as the
program runs. These, `dynamic' (i.e., `dynamically-created') effects
let us, for example, implement
reference cells as instances of the |state| effect. The realization of
this elegant idea required extending Eff with default handlers, with
their special syntax and semantics. The complexity was the reason
dynamic effects were removed from Eff 4.0 (but may be coming back).

The OCaml embedding of Eff gave us the vantage point of view to
realize that dynamic effects may be treated
themselves as an effect. This |New| effect may create arbitrarily many
instances of arbitrary effects of arbitrary types. Below we briefly
describe the challenge of dynamic effects and its resolution in OCaml.

We take the |state| effect as the new running example:
\begin{code}
type 'a state =
  | Get of unit * ('a    -> 'a state result)
  | Put of 'a    * (unit -> 'a state result)
\end{code}
Having defined |get| and |put| effect-sending functions like
|choose| before:
\begin{code}
let get p arg = shift0 p (fun k -> Eff (Get (arg,k)))
let put p arg = shift0 p (fun k -> Eff (Put (arg,k)))
\end{code}
we can use |state| as we did |nondet|. First, however, we abstract the
state handling code into
\begin{code}
let handle_ref s p thunk =
  handle_it p thunk
   (fun v -> fun _ -> v)
   (fun loop -> function
     | Get (_,k) -> fun s -> loop (k s) s
     | Put (s,k) -> fun _ -> loop (k ()) s)
   s
\end{code}
that takes the state effect instance |p|, the initial state |s| and
the |thunk| and handles its |Get| and |Put| requests until it is done.
The handler implements the familiar state-passing technique
\cite{launchbury-state}. Here is a
simple example of using it:
\begin{code}
let a = new_prompt () in              (* instantiate *)
handle_ref 10 a
(fun () ->
  let u = get a () in
  let v = get a () in
  put a (v + 30);
  let w = get a () in
  (u,v,w))
\end{code}
whose result is |(10,10,40)|.

To really treat an instance of |state| as a reference cell, we need a
way to create many |state| effects of many types.
Whenever we need a new reference cell, we should be able to create a
new instance of the |state| signature \emph{and} to wrap the program with
the handler for the just created instance. The first part is easy,
especially in the OCaml embedding: the effect-instance--creating
|new_prompt| is the ordinary function, and hence can be called
anywhere and many times. To just as dynamically put
|handle_ref p s0 ... | around the whole program is complicated. Eff had
to introduce `default handlers' for a signature
instance, with special syntax and semantics. An effect not
handled by an ordinary (local) handler is given to the default
handler, if any.

Our OCaml embedding demonstrates that dynamic effects require nothing
special: Creating a new instance and handling it may be treated as an
ordinary effect:
\begin{code}
type 'e handler_t = {h: 'w. 'e result prompt -> (unit -> 'w) -> 'w}
type dyn_instance =
    New : 'e handler_t * ('e result prompt -> dyn_instance result) -> dyn_instance
let new_instance p arg = shift0 p (fun k -> Eff (New (arg,k)))
\end{code}
The |New| effect receives as the argument the handling function |h|.
The |New| handler creates a new instance |p| and
passes it as the reply to the continuation~-- at the same time
wrapping the handler |h| around the continuation:
\begin{code}
let new_handler p thunk =
  handle_it p thunk
    (fun v -> v)
    (fun loop -> function New ({h=h},k) ->
      let new_instance_p = new_prompt () in
      h new_instance_p (fun () -> loop @@ k new_instance_p))
\end{code}
Both steps of the dynamic effect creation are hence accomplished by
the ordinary handler. The allocation of a reference cell is hence
\begin{code}
let pnew = new_prompt ()
let newref s0 = new_instance pnew {h = handle_ref s0}
###   val newref : 'a -> 'a state result prompt = <fun>
\end{code}
Being polymorphic, |newref| may allocate cells of arbitrary
types. The following is a simple example of reference
cells as |state| instances, with two reference cells |a| and |b|
of two different types:
\begin{code}
let pnew = new_prompt () in
new_handler pnew
(fun () ->
  let newref s0 = new_instance pnew ({h = fun p th -> handle_ref s0 p th}) in
  let a = newref 10 in
  let u = get a () in
  let v = get a () in
  put a (v + 30);
  let b = newref "a" in
  let w = get a () in
  (u,v,w,get b ())
\end{code}
The example yields |(10,10,40,"a")|.

The |New| effect, albeit  `higher-order', is not special.  Programmers
may  write their  own handlers  for it, e.g., to implement
transactional state.

It goes without saying that if a computation uses the |New| effect, it
has to be performed within the scope of the corresponding handler.
That is why the code of the previous example had |new_handler| wrapped
around it. In Eff, the default handlers associated with resources such
as reference cells have global scope and require no `wrapping around'.
To get the similar behavior in OCaml, we have to assume
that the whole program is implicitly wrapped into the |New| effect
handler. One may disagree about infelicity or importance of this
assumption. We only remark that such implicit wrapping is not without
precedent: in OCaml, a program is always wrapped into the default exception
handler, which handles any exception by printing it and terminating
the program.

\section{Evaluation}
\label{s:eval}

In this section, we evaluate the performance for Eff 3.1 embedded in
OCaml (described in \S\ref{s:eff-delimcc})
and compare it against the performance of Eff 3.1, compiled with the
optimizing backend. For the embedded versions, we consider both the delimcc and
the multicore OCaml backends. For the sake of comparison, we
also evaluate the performance of the equivalent program written in \emph{pure}
OCaml, that is, without the use of effects and handlers.

\subsection{N-queens benchmark}

The benchmark we consider is the N-queens benchmark. The aim of the benchmark
is to place N queens on a board of size N such that no two queens threaten each
other. The algorithm involves a backtracking depth-first search for the desired
configuration. For this benchmark, we consider the following 6 versions of the
N-queens program:
\begin{itemize}
\item |Exception|: A pure version with backtracking implemented using native
	OCaml exceptions.
\item |Option|: A pure version with backtracking implemented using an
  option type.
\item |Eff|: An impure version of the benchmark compiled using Eff's optimizing
	compiler backend and with backtracking via
        effect handlers.
\item |Multicore|: An impure version where backtracking is implemented with
	native effects in multicore OCaml.
\item |Eff_of_multicore|: An impure version of the benchmark implemented in Eff
	embedded in OCaml using multicore OCaml handlers.
\item |Eff_of_delimcc|: An impure version of the benchmark implemented in Eff
	embedded in OCaml using the delimcc backend.
\end{itemize}

\begin{figure}[!htpb]
\begin{code}
exception Failure

let main n =
  let l = ref [] in
  for i = n downto 1 do
    l := i::!l;
  done;
  let rec place x qs =
    if x = n+1 then qs else
      let yl = available x qs !l in
      let rec loop = function
        | [] ->
            raise Failure
        | y::ys ->
            try place (x+1) ((x,y) :: qs) with
            | Failure -> loop ys
      in loop yl
  in
  match place 1 [] with
  | res -> print_endline "Success!"
  | exception Failure -> print_endline "Fail: no valid assignment"
\end{code}
\caption{Backtracking N-queens benchmark implemented using exceptions.}
\label{fig:queens_exception}
\end{figure}

The code for the |Exception| version is presented in
Fig.~\ref{fig:queens_exception}, using the auxiliary functions
\begin{code}
let no_attack (x,y) (x',y') = x <> x' && y <> y' && abs (x-x') <> abs (y-y')
let available x qs l          = List.filter (fun y -> List.for_all (no_attack (x,y)) qs) l
\end{code}
Here, |no_attack| returns true if two queens
on the board do not threaten each other. The |available| function, given |qs|,
a safe assignment of queens in the first |x-1| rows, returns the list of
possible safe positions for a queen on the |x|th row. The function
|place| in Fig.~\ref{fig:queens_exception}
attempts to safely place |n| queens, one on each row in a non-threatening
configuration on the board of size |n|. This is done by exploring the possible
assignments in a depth-first fashion. If the search along a path is not
successful, the control backtracks by raising |Failure|, and the next path is
attempted. If successful, the function returns the configuration. The main
function prints a success message if some safe configuration is possible.
Otherwise, it prints an error message.

\begin{figure}[!htpb]
\begin{code}
effect Select : 'a list -> 'a

let queens_multicore n =
  try
    let l = ref [] in
    for i = n downto 1 do
      l := i::!l;
    done;
    let rec place x qs =
      if x = n+1 then Some qs else
        let y = perform @@ Select (available x qs !l) in
        place (x+1) ((x, y) :: qs)
    in place 1 []
  with
  | effect (Select lst) k ->
      let rec loop = function
        | [] -> None
        | x::xs ->
            match continue (Obj.clone_continuation k) x with
            | None   -> loop xs
            | Some x -> Some x
      in loop lst
\end{code}
\caption{Backtracking N-queens benchmark implemented using multicore OCaml effect handlers.}
\label{fig:queens_multicore}
\end{figure}

Fig.~\ref{fig:queens_multicore} shows the |Multicore| version of the N-queens
benchmark. We declare an effect `Select' which is parameterized with a list of
elements of some type, which when performed returns an element of that type.
For placing each queen, in the |place| function, we perform the effect `Select'
with the list of available positions for the next queen. The effect handler
performs backtracking search and explores each of the possibilities by invoking
the continuation with different assignments for the position of the next queen.
Since continuations in multicore OCaml are one-shot by default, we need to
clone the continuation before we resume the continuation. The cost of cloning
is linear in the size of the continuation.

\subsection{Results}

\begin{figure*}[htpb]
  \includegraphics[width=\textwidth]{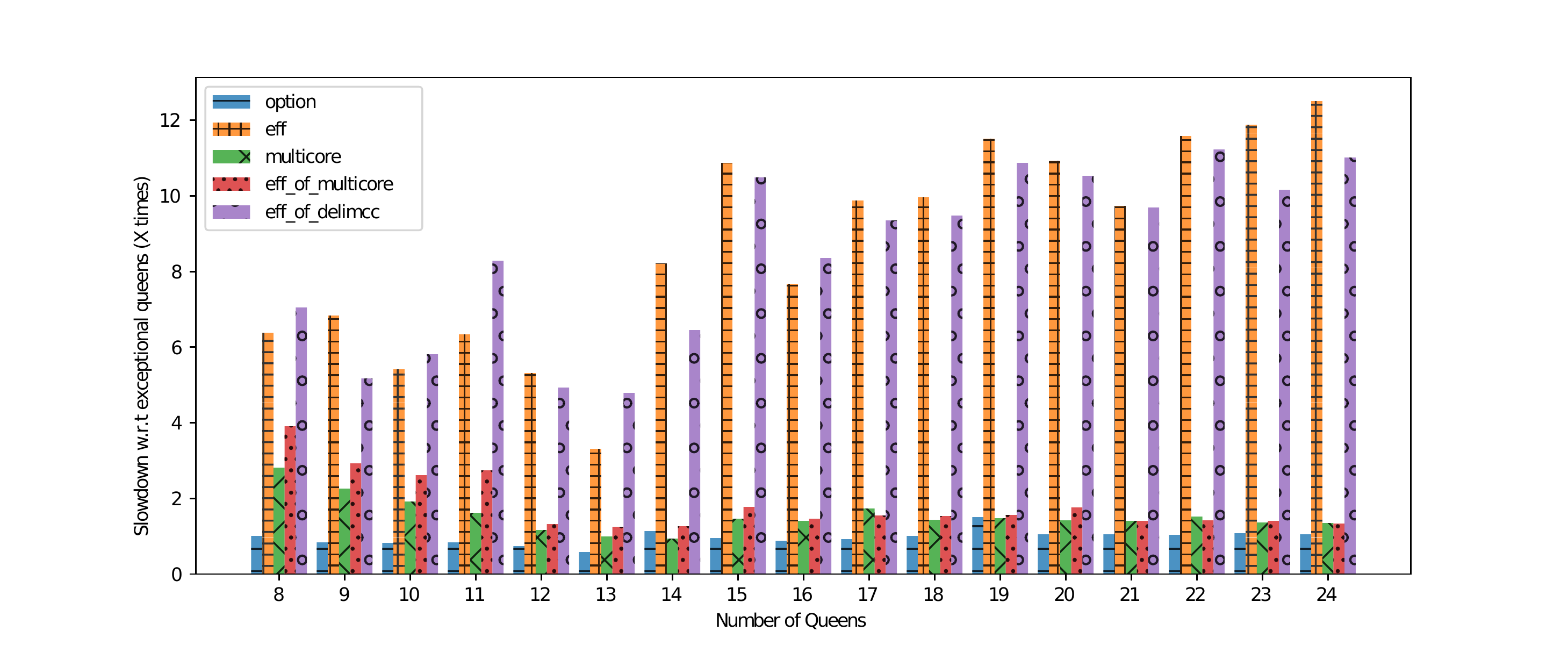}
  \caption{Performance comparison on N-queens benchmark.}
  \label{fig:queens_perf}
\end{figure*}

Fig.~\ref{fig:queens_perf} shows the performance of different versions of the
N-queens benchmark. The experiments were performed on an 2016 MacBook Pro with
3 GHz Intel Core i7 processor and 16 GB of DDR3 main memory. The machine had 2
cores and 4 hardware threads and was unloaded at the time of experiments.

The results show the running times of each version normalized to the
|Exception| version, as we increase the size of the board. The results show
that the pure OCaml versions perform best and on par with each other. This is
unsurprising since these versions do not incur the cost of effect handlers and
reifying the continuations. The |Multicore| version performs best among the
effectful versions. Multicore OCaml implements effect handlers natively with
the help of first-class runtime support for delimited continuations that is
fully integrated into OCaml's runtime system. As a result, installing effect
handlers and continuation capture are cheap operations. We observed that Eff
embedding in Multicore OCaml was only $1.2\times$ slower than the exception
version on average. |Eff_of_multicore| performs marginally slower than
|Multicore| due to boxing overheads.

\begin{table*}
	\begin{center}
	\begin{tabular}{| l | r |}
		\hline
		Configuration & Allocation (GB) \\
		\hline
		|Exception| 				& 0.62 \\
		|Option| 						& 0.62 \\
		|Multicore| 				& 0.82 \\
		|Eff_of_multicore|  & 1.11 \\
		|Eff_of_delimcc|		& 0.88 \\
		|Eff|								& 44.71 \\
		\hline
	\end{tabular}
\end{center}
\caption{Total memory allocated for the different N-queens program
  configurations for a board size of 24, over multiple GC cycles
  during the lifetime of the program. The maximum resident set size as 
  reported by GNU time command is around 5MB for all configurations.}
\label{tab:memory}
\end{table*}

The |Eff| version and |Eff_of_delimcc| are comparatively slower than the other
versions. This is because delimcc is designed to be an independent library that
requires no change to the OCaml compiler and the runtime. The cost of this
generality is that delimcc continuation capture and management are more
expensive than continuations in multicore OCaml. On average, the
|Eff_of_delimcc| version is $8.5\times$ slower than the exception version.
However, both the embeddings, |Eff_of_delimcc| and |Eff_of_multicore| are
faster than native, optimized Eff versions. The Eff implementation of handlers
is through Free monadic interpretation, incurring the cost of intermediate
closures even for pure OCaml code. While the Eff compiler optimizes primitive
operations, there are large overheads from the use of the Free monad for the
rest of the program. This can clearly be seen in the results presented in
Table~\ref{tab:memory}. On average, the Eff version is $9.9\times$ slower than
the exception version.

\section{Related Work}
\label{s:related}

The key insight underlying various implementations of effects is
treating an effectful operation as `sending a mail' to the `authority'
(handler, or interpreter). The mail has the message (effect
parameters) and the return address, represented as a delimited
continuation. An interpreter examines the mail message and may
send the reply, upon receiving which the original computation resumes.
This insight appears already in the very first paper on delimited
control \cite{felleisen-prompt} and was fully developed in
\cite{cartwright-extensible}. The handlers for effect messages do not
have to be all at the `top level', they can be distributed throughout
the program. Such a refinement was first used in
\cite{remove-dyn-prompt} to prove that all variations of the `shift'
operator (shift itself, shift0, control, control0) are equally
expressible, in the untyped setting. That approach has later led to
extensible effects \cite{exteff,freer}. Embedding Eff in OCaml may be
seen as porting of extensible-effects to OCaml, with delimited control
operators instead of continuation-passing style, and the `out-of-band'
emulation of answer-type polymorphism.

Algebraic effects in OCaml were also implemented in 
Kammar et al.~\cite{kammar-handlersinaction}, also in terms of the 
delimited control
operator shift0. However, Kammar's encoding relies on the global
mutable variable holding the stack of handlers in the current dynamic
scope. Global mutable cells preclude a `local' (i.e., macro)
translation from Eff to OCaml and complicate reasoning.

Recently Forster et al.~\cite{ForsterKLP16} presented the encoding of
a simple Eff calculus into a delimited control calculus that is close
to ours in spirit. The authors relied on a very different formalism
of an extended call-by-push-value. Their Eff calculus was also bigger,
compared to our single-operation Core Eff. The
correctness proof was given operationally: the delimited control
calculus simulates the Eff calculus up to congruence. The main difference
from our work (beside the operational vs. denotational distinction) is
that the encoding of Forster et al. does not preserve typeability: not
surprisingly because of the answer-type polymorphism (which the authors
could neither represent nor emulate in their system).

Our denotational semantics of Core Eff and Core delimcc are expressed
in the tagless-final style and take the form of an
interpreter. Definitional interpreters and their defunctionalized
versions (abstract machines) for delimited control are well-known:
\cite[Fig.1]{biernacka-operational} for the ordinary |shift|, and
\cite[Fig.1]{dybvig-monadic} for the multi-prompt delimited control. These
machines and interpreters work with untyped source language. They
are written to evaluate programs that include delimited
control; it is rather hard to see from them what the meaning of
|shift| by itself is. After some eyestrain one sees
the continuation semantics of |shift| and multi-prompt |shift|
\cite{biernacka-operational,dybvig-monadic}, which does tell the meaning of
the mere |shift|, compositionally~-- and hence may be regarded as
denotational. The difference of our denotational semantics is
the formulation without resorting
to continuation-passing style and without continuation stacks,
meta-continuations, etc. The so-called direct-style of our semantics
seems to make the reasoning simpler.

\section{Conclusions and the Further Research Program}

We have demonstrated the embedding of Eff 3.1 in OCaml by a simple,
local translation, taking advantage of the delimcc library of
delimited control. We may almost copy-and-paste Eff code into OCaml,
with simple adjustments. The embedding not only lets us play with Eff
and algebraic effects in \emph{ordinary} OCaml. (Recall, that
multicore OCaml is still an unofficial dialect.) It also
clarified the thorny dynamic effects, demonstrating that there is
nothing special about them. The delimited control turned out very helpful in
quickly prototyping dynamic effect handling and reaching that
conclusion. Once it is
realized that dynamic effect creation can be treated as
an ordinary effect, dynamic effects can now be supported in multicore
OCaml and other effect frameworks. The OCaml embedding has inspired
other Eff embeddings, such as the one into F\# by Nick
Palladinos\footnote{\url{http://github.com/palladin/Eff}}.

An unexpected conclusion is that the seemingly well-researched area
of delimited control still harbors hidden vistas. First is the
direct denotational semantics of delimited control. We have just seen how
useful the denotational approach has been, in proving the correctness
of the translation from Eff to OCaml. It seems worthwhile to consider
the denotational semantics for multicore OCaml, relating it directly
to Eff.

The occurrences of |Obj.magic| and of the universal type have surely
caught the eye. Are such
concessions inevitable if one stays with relatively simple types?
Or are they merely an artifact of an inadequate interface of delimited
control? Following the well-established analogy between
control operators and exceptions, one may see that |push_prompt| (also
called reset) corresponds to the following rather
specific exception-catching form: |try expr with exc -> exc|.
Although there are indeed cases for which such a limited form of
exception-catching is appropriate, most of the time we wish to
distinguish the normal and the exceptional termination of the expression
|expr|. Likewise we wish to distinguish the normal and the shiftful
termination of |expr| in |push_prompt p expr|, and hence
have to work around the restricted interface of |push_prompt| by
defining the sum data type such as |free|. One wonders
if a better interface for delimited control can be designed, without
unnecessary restrictions and with simpler typing rules.

Finally, it is interesting to see how higher-order (dynamic) effects
can be expressed in a type-and-effect system, where the type of an
expression tells not only its result but also the effects it may
execute.

\subsubsection*{Acknowledgments}
We are very grateful to Andrej Bauer for introducing us to Eff, for patiently
explaining Eff features and design decisions, and for writing some of the
sample Eff code in \S\ref{s:eff-intro}. We thank Kenichi Asai,
Yukiyoshi Kameyama and Achim Jung for helpful
discussions. Extensive comments and suggestions by anonymous reviewers
are greatly appreciated.
This work was partially supported by JSPS KAKENHI Grant Number
17K00091.

\bibliographystyle{eptcs}
\bibliography{refs.bib}
\end{document}